\documentclass[fleqn,usenatbib]{mnras}
\usepackage{amsmath}
\usepackage{amssymb}
\usepackage{graphicx}
\usepackage{float}
\usepackage[T1]{fontenc}
\usepackage{multirow}

\usepackage{xcolor}
\colorlet{purple}{blue!50!red}
\colorlet{orange}{red!50!yellow}

\newcommand{\add}[1]{\textcolor{black}{#1}}

\begin{document}

\title[Subhalo density slopes from HST strong lensing data with likelihood-ratio estimation]{Subhalo effective density slope measurements from HST strong lensing data with neural likelihood-ratio estimation}
\author[G. Zhang, A. Ç. Şengül, and C. Dvorkin]{Gemma Zhang$^{1}$\thanks{\href{mailto:yzhang7@g.harvard.edu}{yzhang7@g.harvard.edu}}, Atınç Çağan Şengül$^{1}$, and Cora Dvorkin$^{1}$
\\\\
$^{1}$Department of Physics, Harvard University, Cambridge, MA 02138, USA\\
}

\label{firstpage}
\pagerange{\pageref{firstpage}--\pageref{lastpage}}
\maketitle

\begin{abstract}
    Examining the properties of subhalos with strong gravitational lensing images can shed light on the nature of dark matter. From upcoming large-scale surveys, we expect to discover orders of magnitude more strong lens systems that can be used for subhalo studies. To optimally extract information from a large number of strong lensing images, machine learning provides promising avenues for efficient analysis that is unachievable with traditional analysis methods, but application of machine learning techniques to real observations is still limited. We build upon previous work, which uses a neural likelihood-ratio estimator, to constrain the effective density slopes of subhalos and demonstrate the feasibility of this method on real strong lensing observations. To do this, we implement significant improvements to the forward simulation pipeline and undertake careful model evaluation using simulated images. Ultimately, we use our trained model to predict the effective subhalo density slope from combining a set of strong lensing images taken by the \textit{Hubble Space Telescope}. We found the subhalo slope measurement of this set of observations to be steeper than the slope predictions of cold dark matter subhalos. Our result adds to several previous works that also measured high subhalo slopes in observations. Although a possible explanation for this is that subhalos with steeper slopes are easier to detect due to selection effects and thus contribute to statistical bias, our result nevertheless points to the need for careful analysis of more strong lensing observations from future surveys.
\end{abstract}

\begin{keywords}
gravitational lensing: strong -- dark matter
\end{keywords}

\section{Introduction}

The standard $\Lambda$CDM cosmological model has been in remarkable agreement with large-scale observations, but there is scarce evidence for the nature of dark matter on small (sub-galactic) scales. Because the nature of dark matter (DM) remains elusive, examining various dark matter models using small-scale cosmological observables becomes crucial. One of the promising observables used to study DM is subhalos, which are small dark matter clumps gravitationally bound to a larger halo. Probing the properties of these subhalos can potentially shine light on the nature of DM, as subhalos exhibit different properties under alternate DM models beyond cold dark matter (CDM). For instance, warm dark matter (WDM) models predict a smaller number of low-mass subhalos and more cored subhalo density profiles compared to CDM~\citep{bode2001wdm}, while self-interacting dark matter (SIDM) models generally predict more cored subhalo profiles than that of the CDM model~\citep{spergel2000sidm}. 

Because low-mass subhalos are observed to lack luminous matter~~\citep{fitts2017fire, read2017stellar, kim2018missing}, they are typically probed through their gravitational effects. Strong gravitational lensing, a predicted phenomenon from General Relativity, is a powerful way to constrain subhalo properties. In strong gravitational lensing, light emitted by a distant source gets deflected by the gravitational field of a massive structure (lens), and properties of the lens and its substructure can be inferred by analyzing the images of the source light. In this paper, we will focus on studying subhalos in the lens galaxy of strong lensing systems in which both the lens and background source are galaxies. 

To date, there have been a few claimed detections of substructure in galaxy-galaxy strong lensing observations~\citep{vegetti2010detection, vegetti2012detection, hezaveh2016detection}. Existing analyses that use observed strong lensing images to constrain DM models primarily rely on modeling individual (often the most massive) substructure in a lens system~\citep{vegetti2014inferencecdmsubmf, ritondale2019lowmass, sengul2022reanalyzed}. While useful, direct substructure modeling is computationally costly, and it is often limited to capturing the effect of relatively massive subhalos. Even though the CDM paradigm predicts a large number of subhalos with smaller masses, they are difficult to probe through traditional analysis methods because the inclusion of more subhalos makes sampling of the joint parameter space prohibitive. As a result, it is important to explore alternative analysis methods that can more optimally incorporate information from the large population of smaller subhalos. 

To leverage the collective effect of subhalo populations on strong lensing images, there has been significant work done to obtain statistical constraints from subhalos~\citep{dalal2001direction, hezaveh2014measuring, cyr-Racine2015darkcensus, diazRivero2017powerspec, birrer2017lensingsubs, diazRivero2018powerspecethos, daylan2018probing, brennan2018quantifying, gilman2018, cyrracine2019, he2022}. In particular, machine learning has emerged as a promising candidate to analyze subhalos in strong lensing images for its ability to efficiently and implicitly marginalize over a large parameter space. With the upcoming large-scale imaging surveys, the number of observed strong lensing systems is expected to increase significantly~\citep{laureijs2011euclid,Collett:2015roa,mckean2015ska, bechtol2019lsst, jacobs2019des, huang2021desi, storfer2022desi}. Machine learning has a much-needed advantage that can make inference on these large datasets feasible. 

Several deep learning techniques have been demonstrated to be effective at constraining the subhalo mass function using simulated strong lensing images~\citep{brewer2016transdim, brehmer2019mining, Ostdiek:2020mvo, Ostdiek:2020cqz, 2022arXiv220509126A, wagner2022images}, but so far, there has been no successful attempt at applying them to real observations. The main challenge of applying deep learning methods to observations comes from the need for the training set to closely resemble the test set, as deep learning models are known to struggle in the presence of a distribution shift between training and test sets~\citep{recht2018, recht2019}. Most of the previous works on machine learning applications to strong lensing made simplifying assumptions in the forward modeling pipeline of the training set in order to demonstrate the potential suitability of a method. However, for the machine learning model to be deployed on observations, its training set needs to incorporate all possible complexities that exist in the observed data. 

In this work, for the first time, we analyze subhalo properties in real strong lensing observations with a machine learning technique. We build upon the method developed in~\cite{zhang2022} by adding multiple layers of complexities in the forward pipeline for the training set. Through training, our model learns to infer the effective subhalo density slope (directly related to the subhalo concentration), a promising observable proposed by~\cite{sengul2022probing} for distinguishing DM models. Several other works have also shown that the concentration of subhalos is an effective probe of DM~\citep{minor2021inferring, minor2021unexpected, amorisco2022}. Using our trained model on real observed strong lensing images, we found a subhalo density slope steeper than those of subhalos predicted by the CDM model. This measurement is consistent with previous works, which also found unexpectedly large subhalo concentrations~\citep{minor2021unexpected, sengul2022probing}.

This paper is organized as follows. In Sec.~\ref{sec:data}, we discuss details of the forward model used to generate mock strong lensing images. In Sec.~\ref{sec:model}, we summarize the deep learning technique that we use for inference, discuss our inference procedure, and outline our neural network architecture. In Sec.~\ref{sec:results}, we evaluate our trained model and compare the model predictions on the observed data with those under the CDM model. We conclude with a summary of our results in Sec.~\ref{sec:conclusion}, and discuss the implications of our work.

\section{Data}
\label{sec:data}
 
We generate simulated lensing images to train our neural network and compare our model predictions with ground truths on mock images post-training to ensure training quality. At inference time, we apply the trained model to a set of observed lensing images from the {\it Hubble Space Telescope} (HST). We discuss details of both the mock data and the real HST observations below. 

\subsection{Mock data generation}

To generate our mock strong lensing images, we use the software package \textsc{lenstronomy}~\citep{birrer2015gravitational, birrer2018lenstronomy}. In order to match the HST post-drizzling image configuration, we generate ($100 \times 100$)~pixel$^2$ images, with a resolution of $0.04''$ per pixel. We build upon the simulation pipeline used in~\cite{zhang2022} and include significantly more complexities in the modeling process so as to make the images as similar to real observations as possible. Modeling a strong lens system requires several ingredients in the forward model: a source galaxy, a main (host) lens galaxy, a population of subhalos and light-of-sight (LoS) halos. In addition, we specify the instrumental configuration and image pre-processing of the mock images in the forward simulation. The distributions of parameters governing the lens models of our simulated images are summarized in Table~\ref{table:params}. 

\subsubsection{Source and main lens}

In a galaxy-galaxy lens system, light rays of a background source galaxy get gravitationally deflected by a foreground lens galaxy en route to the detector. Strong gravitational lensing specifically refers to the case where the projected surface mass density of the lens is greater than the critical surface density $\Sigma_{\mathrm{crit}}$. In this scenario, the bending of source light is significant enough to result in characteristic arcs of light in observed images.

To simulate the source light, we use galaxy images taken by the HST Cosmic Evolution Survey~(COSMOS)~\citep{scoville2007cosmos, koekemoer2007cosmos} processed by \textsc{paltas}~\citep{wagner2022images}. The \textsc{paltas} package takes a sub-sample of the HST COSMOS survey galaxy images~\citep{mandelbaum2012precision, mandelbaum2014great3} and filters out suitable source candidates. To simulate the source for each mock image, we randomly draw a galaxy image from the COSMOS catalog and randomly vary the rotation angle and the source coordinates $(x_{\mathrm{source}}, y_{\mathrm{source}})$. From the 2,262 available source galaxies, we use 2,163 (96\%) for the training set, 70 (3\%) for the validation set, and the remainder (1\%) for testing and evaluation. 

We model the main-lens mass distribution using an elliptical power law~(EPL) profile~\citep{barkana1998epl}. The convergence of an EPL profile at position $(x, y)$ on the lens plane is given as follows: 
\begin{align}
    \kappa(x, y) = \frac{3 - \gamma}{2} \left(\frac{\theta_E}{\sqrt{qx_{\phi}^2 + y_{\phi}^2/q}}\right)^{\gamma - 1},
    \label{eq:epl_kappa}
\end{align}
where $\theta_E$ is the Einstein radius, $q$ is the minor/major axis ratio, $x_{\phi}, y_{\phi}$ are positions on the axes aligned with the major and minor axes, and $\gamma$ is the power-law slope of the mass distribution. To model each main lens, we draw its $\gamma_{\mathrm{ML}}$ ($\gamma$ of the main lens) from $\mathcal{N}(2, 0.1)$ and truncate the tails of the normal distribution so that the range of possible values is bounded by 1.1 and 2.9. Slope values outside of the $(1, 3)$ interval lead to nonphysical or divergent mass profiles and are thus not included in our modeling. Adding variations in $\gamma_{\mathrm{ML}}$ simulates the natural stochasticity in lens density profiles that deviate from an isothermal profile ($\gamma = 2$). Note that $\phi$ indicates the angle between the major/minor axes and the fixed $(x, y)$ axes of an image. The inputs into \textsc{lenstronomy} are the ellipticity moduli, which are directly related to $q$ and $\phi$: 
\begin{align}
    e_1 &= \frac{1 - q}{1 + q} \cos(2\phi), \\ 
    e_2 &= \frac{1 - q}{1 + q} \sin(2\phi).
\end{align}
In addition, we add multipole moments $m = 3, 4$ to the EPL lens mass distribution in order to more realistically model the mass distribution of more complex lenses that may deviate from an elliptical profile. We also include an external shear parametrized by $\gamma_{\mathrm{shear}, 1}$ and $\gamma_{\mathrm{shear}, 2}$~\citep{Keeton1996shear}. The shear parameters $\gamma_{\mathrm{shear}, 1}$ and $\gamma_{\mathrm{shear}, 2}$ are the diagonal and off-diagonal terms of the shear matrix, respectively. 

In~\cite{zhang2022}, it is assumed that the light produced by the lens galaxy has already been subtracted from the original observed image through a coarse modeling process. However, for real observed images, removing the lens light may involve imperfect modeling and high computational cost. To bypass this assumption, we include lens light in our mock image modeling. We assume that the center of the lens light coincides with the center of its mass density profile and that the lens light takes on an elliptical S\'ersic profile~\citep{sersic1963}, with the brightness parametrized as:
\begin{align}
    I(r) = I_0 \mathrm{exp}\left(-b_{n_{\mathrm{sersic}}}\left(\frac{r}{r_{\mathrm{sersic}}}\right)^{\frac{1}{n_{\mathrm{sersic}}}}\right), 
\end{align}
where $n_{\mathrm{sersic}}$ is the S\'ersic index, $b_{n_{\mathrm{sersic}}} \approx 1.999 n_{\mathrm{sersic}} - 0.327$. Here, $r_{\mathrm{sersic}}$ is the half-light radius, and $I_0$ is determined by the apparent magnitude~\citep{birrer2018lenstronomy}. We draw the apparent magnitude of the lens light from a uniform distribution between 17 and 19. We choose this range to be consistent with the apparent magnitude measurements of lens galaxies in the observed images used in our analysis~\citep{auger2009slacs}, which are discussed in detail in Sec.~\ref{subsec:hst_data}. In each mock image, we vary all parameters governing the lens model, including its center position, Einstein radius, shear parameters, apparent magnitude, and its ellipticity. The variation ranges of these parameters are summarized in Table~\ref{table:params}. 

In simulating our training set images, we take into account the spectroscopic redshifts of the real HST observations used during inference. To simulate each image in our training set, the source galaxy redshift is drawn from a uniform distribution of $z_{\mathrm{source}} \sim \mathcal{U}(0.5, 0.7)$, while the lens galaxy redshift is drawn from a uniform distribution of $z_{\mathrm{lens}} \sim \mathcal{U}(0.15, 0.25)$. These redshift ranges roughly match with those of the real observations that we use for inference. We deliberately chose to work with systems with relatively low source redshifts because they align better with the redshifts of the COSMOS galaxies that are used in our modeling pipeline, minimizing the difference between our simulated images and the real observations. 

\subsubsection{Subhalos and line-of-sight halos} 
\label{subsubsec:subhalo_and_los}

Aside from the main lens, the observed strong lensing images of the source light are affected by additional structures: subhalos, which are small halos residing inside the main host halo, and line-of-sight (LoS) halos, which are located along the line-of-sight between the source galaxy and the observer. If these (sub)halos are found within the bright lensed arcs in the observed image, they can leave detectable perturbations on the observed images. Analyzing these perturbations provides us with information about the properties of these substructures. 

In our pipeline for simulating the training and validation set images, we add subhalos and LoS halos that follow the EPL profile given in Eq.~\ref{eq:epl_kappa}. The $\gamma$ parameter in the EPL profile controls the steepness of the halo density profiles: a larger $\gamma$ implies a denser halo density profile. We model our training and validation set images with EPL (sub)halos because it allows us to label each image with its underlying power-law slope, which is the ultimate parameter of interest during our inference. We include a uniform prior on $\gamma$ in our training set so that we do not unnecessarily bias our model. To model the subhalos and LoS halos in each image, we first draw $\gamma$ from a uniform distribution: $\gamma \sim \mathcal{U} (1.1, 2.9)$; we then draw normally distributed slopes $\gamma_i \sim \mathcal{N}(\gamma, 0.1\gamma)$ for the $i$th subhalo. This normal distribution is truncated so that $\gamma_i$ is constrained between 1.01 and 2.99. The number of subhalos added to each image is drawn from a uniform distribution $N_{\mathrm{sub}} \sim \mathcal{U}\{0, 3000\}$. Note that the upper bound of $3000$ subhalos is an overestimate of a realistic number of subhalos for our host halo mass, but we include a higher number of subhalos so that our neural network can successfully learn the signatures in the lensed images corresponding to the changes in the density slope $\gamma$. During model evaluation, we will use a smaller $N_{\mathrm{sub}}$ range to simulate a more realistic substructure fraction, as will be discussed in Sec.~\ref{sec:results}. 

Because only subhalos near the bright arcs of an image have observable effects, in our simulated images, we place subhalos solely in pixels whose brightness is more than a fifth of the maximum brightness in the smooth model image, which is the image modeled with only the lens and source galaxies and no substructure (and in this case no lens light). The Einstein radius of each subhalo is determined by its mass $M_{200}$, which is drawn from a subhalo mass function $dN_{\mathrm{sub}}/dM_{200} \propto M_{200}^{-1.9}$. The mass $M_{200}$ is defined as the total mass enclosed by $r_{200}$, which is the radius within which the average mass density is 200 times the critical density of the Universe. In our simulated images, we only add subhalos with masses between $10^7\mathrm{M}_{\odot}$ and $10^{10}\mathrm{M}_{\odot}$, because subhalos heavier than this range are scarce and can often be individually modeled. 

To add the LoS halos in our modeling, we use the pipeline provided by \textsc{paltas}, with several added modifications. The properties of each LoS halo are determined by the following parameters: its mass $M_{200}$, density slope $\gamma_i$, ellipticities $e_1, e_2$, redshift $z_{\mathrm{los}}$, and position coordinates $x_{\mathrm{los}}, y_{\mathrm{los}}$. \textsc{paltas} determines the mass $M_{200}$ of each LoS halo using a modified Sheth-Tormen halo mass function, which includes two additional free parameters to the mass function proposed originally in~\cite{sheth2001}: 1) an overall scaling factor that accounts for uncertainties in the normalization of the mass function; 2) a parameter that accounts for the contribution from the two-point halo correlation function, due to the fact that dark matter halos are biased tracers of the overall matter distribution. The two-point halo correlation function correction is only added for halos along the line-of-sight that are sufficiently close to the main halo of the lens. For a more comprehensive discussion of the modified Sheth-Tormen mass function used in \textsc{paltas}, we refer readers to \cite{wagner2022images}. The density slopes and ellipticities of LoS halos are drawn from the same uniform distributions as subhalos, as discussed above. To determine the position of LoS halos, we divide the space between the observer and the source galaxy into thin slices of redshift with uniform thickness. The position coordinates, ($x_{\mathrm{los}}, y_{\mathrm{los}}$), are bounded by a double cone whose bases lie in the lens plane and whose apexes lie at the observer and the source. The position of each LoS halo is sampled uniformly in the volume of the double cone. 

The addition of subhalos causes an enlarged effective Einstein radius, so to restore the effective Einstein radius to its smooth model counterpart, we add a negative mass sheet in the lens plane for the subhalos. In addition, to avoid making the region along our line-of-sight overdense compared to the rest of the Universe, we add a negative mass sheet in each redshift slice for the LoS halos. The negative mass sheet is a constant sheet of convergence such that the sum over all its pixels cancels out the total convergence added by the subhalos or LoS halos. 

\subsubsection{Instrumentation details and data pre-processing}

To make the mock images as similar to real observations as possible, we incorporate HST instrumentation details in the production of our training set. We model our images using the HST ACS/WFC F814W filter configuration and apply an empirical point spread function (PSF), obtained from examining the exposure of point-like stars~\citep{anderson2000}. We add noise using an exposure time of 2200~seconds, in approximate agreement with the noise level of the observed HST images discussed in Sec.~\ref{subsec:hst_data}. 

Moreover, in real observations, there are often bright structures close to the strong lens system of interest that can potentially distract our analysis. During training, we apply a circular mask to cover the region outside of the lensed arcs, so that our model learns to not get confused by potential confounders. To mask out the edges of the images, we set the area outside of a circular mask to zero after an image has been whitened. We vary the radius of the circular mask based on the Einstein radius of the image.

\begin{table*}
    \centering
    \begin{tabular}{l c} 
        \hline
        \bf{Parameter} & \bf{Distribution} \\
        \hline 
        \multicolumn{2}{l}{\underline{Source}} \\
        Source redshift & $z_{\mathrm{source}} \sim \mathcal{U}(0.5, 0.7)$ \\
        $x$-coordinate & $x_{\mathrm{source}} \sim \mathcal U(-0.1'', 0.1'')$ \\
        $y$-coordinate & $y_{\mathrm{source}} \sim \mathcal U(-0.1'', 0.1'')$ \\
        \hline 
        
        \multicolumn{2}{l}{\underline{Main lens}} \\
        Lens redshift & $z_{\mathrm{lens}} \sim \mathcal{U}(0.15, 0.25)$ \\
        $x$-coordinate & $x_{\mathrm{lens}} \sim \mathcal U(-0.2'', 0.2'')$ \\
        $y$-coordinate & $y_{\mathrm{lens}} \sim \mathcal U(-0.2'', 0.2'')$ \\
        Einstein radius & $\theta_E \sim \mathcal U(0.9'', 1.3'')$ \\
        Ellipticities & $e_1 \sim \mathcal U(-0.2, 0.2) \quad e_2 \sim \mathcal U(-0.2, 0.2)$ \\
        Multipole moments ($m = 3, 4$) & $a_m \sim \mathcal U(0, 0.05) \quad \phi_m \sim \mathcal U(-\pi, \pi)$
        \\
        EPL slope of density profile & $\gamma_{\mathrm{ML}} \sim \mathcal{N}(2, 0.1) $
        \\
        External shear & $\gamma_{\mathrm{shear}, 1} \sim \mathcal U(-0.1, 0.1) \quad \gamma_{\mathrm{shear}, 2} \sim \mathcal U(-0.1, 0.1)$ \\
        
        \hline 
        \multicolumn{2}{l}{\underline{Lens light}} \\
        Apparent magnitude & $m \sim \mathcal U(17, 19)$ \\
        Half light radius & $R_{\mathrm{sersic}} \sim \mathcal N(0.8, 0.15)$ \\
        S\'ersic index & $n_{\mathrm{sersic}} \sim \mathcal N(2, 0.5)$ \\
        Ellipticities & $e_1 \sim \mathcal U(-0.1, 0.1) \quad e_2 \sim \mathcal U(-0.1, 0.1)$\\
        \hline 
        
        \multicolumn{2}{l}{\underline{LoS halos}} \\
        EPL ellipticities & $e_1 \sim \mathcal U(-0.2, 0.2) \quad e_2 \sim \mathcal U(-0.2, 0.2)$ \\
        EPL slope of density profile per lens system & $\gamma \sim \mathcal U(1.1, 2.9)$ \\
        EPL slope of density profile per subhalo & $\gamma_i \sim \mathcal{N}(\gamma, 0.1\gamma)$ \\
        LoS halo mass & $M_{200} \in [10^7, 10^{10}]\mathrm M_{\odot}$ \\
        Halo mass function normalization & $\delta_{\mathrm{los}} \sim \mathcal{U}(0, 2)$ \\
        \hline 
        
        \multicolumn{2}{l}{\underline{Subhalos}} \\
        EPL ellipticities & $e_1 \sim \mathcal U(-0.2, 0.2)$ \quad $e_2 \sim \mathcal U(-0.2, 0.2)$ \\
        EPL slope of density profile per lens system & $\gamma \sim \mathcal U(1.1, 2.9)$ \\
        EPL slope of density profile per subhalo & $\gamma_i \sim \mathcal{N}(\gamma, 0.1\gamma)$ \\
        Subhalo mass function power-law slope & $-1.9$ \\
        Subhalo mass & $M_{200} \in [10^7, 10^{10}]\mathrm M_{\odot}$ \\
        \hline 
    \end{tabular}
    \caption{Parameters of the main components of a galaxy-galaxy strong gravitational lensing system and their respective training distributions in our forward simulation pipeline.}
    \label{table:params}
\end{table*}

\subsection{HST observations}
\label{subsec:hst_data}

We demonstrated, in ~\cite{zhang2022}, that a neural likelihood-ratio estimator is capable of extracting subhalo population density slope information from simulated strong lensing images. In this work, we apply the same method to real strong lensing data taken by the HST. Specifically, we use strong lens systems identified by the Sloan Lens ACS (SLACS) survey~\citep{bolton2008slacs} and followed up by HST observations. 

\begin{figure*}
    \centering
    \includegraphics[width=1\linewidth]{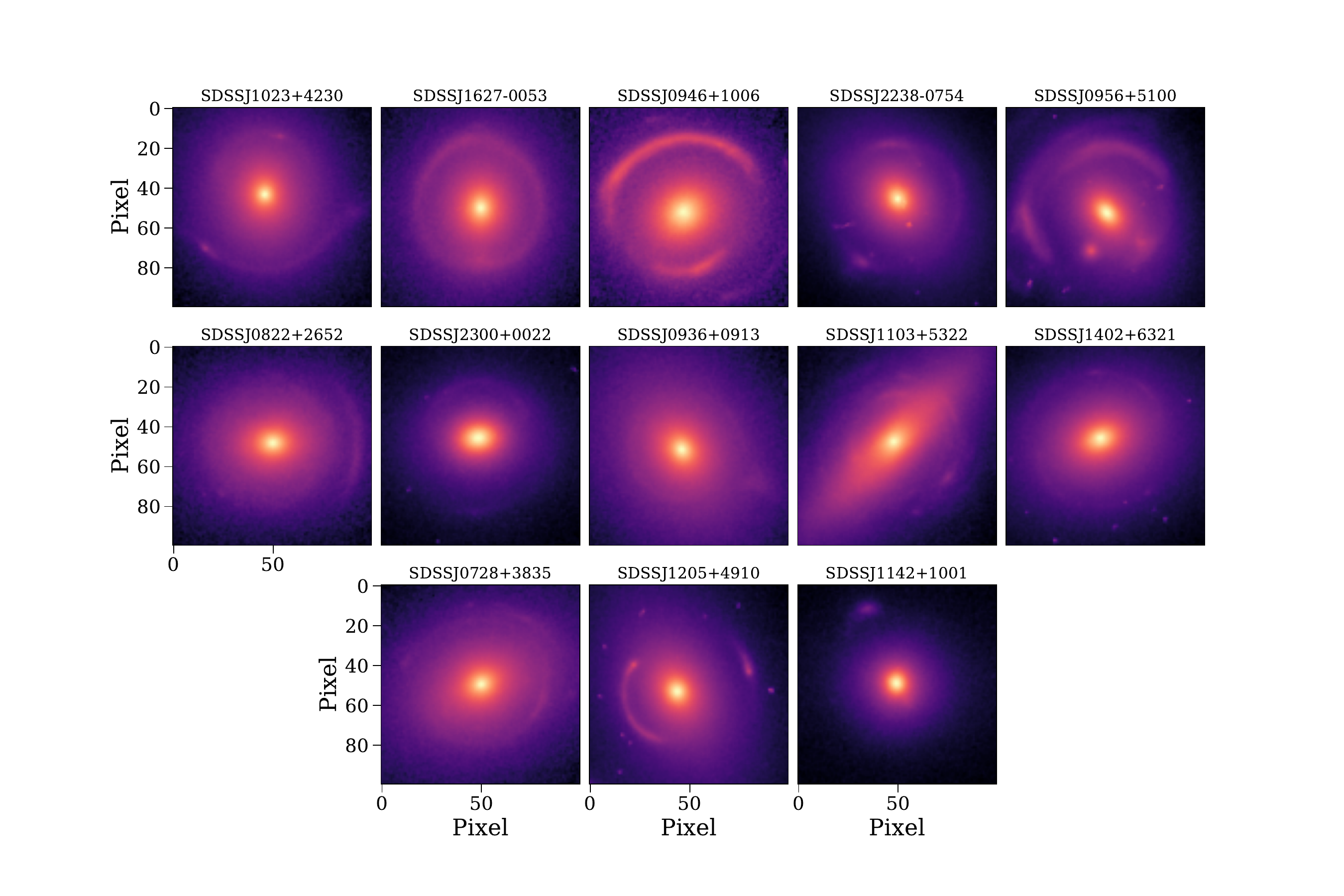}
    \caption{For our inference, we selected 13 image cutouts from observations that share similar instrumental configurations taken by the HST F814W filter. Each image cutout has 100 pixels of size 0.04$''$ per side. The pixel values of the images are shown in log scale so that features of the lensed arcs are visible by eye. The title of each image corresponds to the name of the strong lens system.}
    \label{fig:hst_images}
\end{figure*}

In the SLACS strong lens systems, the redshifts of the foreground galaxies range from 0.05 to 0.5, while the redshifts of the background galaxies range from 0.2 to 1.2. For our analysis, we choose observed images that share the same set of properties, and then simulate a matching training set. The shared properties include $0.04''$ pixel resolution, F814W camera band, and exposure time of approximately 2200 seconds. We also select lens systems with source redshifts, lens redshifts, and Einstein radii that fall in a reasonably narrow range to limit the span of the overall parameter space. From the HST observations, we made $(100 \times 100)$~pixel$^2$ cutouts in which the lens systems of interest are located roughly at the center. Out of these cutouts, we then selected a subset of them that contain visible lensed arcs. The selected HST observations share the same instrumentation details with our training set so as to avoid having an unnecessary distribution shift between training and testing. Our selection process led to a subset of 13 images that we ultimately used for inference, as shown in Fig.~\ref{fig:hst_images}. 

\section{Model and inference}
\label{sec:model}

To infer the subhalo density slopes, we use a simulation-based inference (SBI) machine learning technique. SBI methods have gained increasing popularity in parameter inference problems in cosmology because of their ability to approximate intractable likelihoods due to complicated physical processes. In our application, we train a neural likelihood-ratio estimator as a parametrized classifier to learn the likelihood function~\citep{cranmer2015approximating, baldi2016parameterized,hermans2019likelihoodfree}. 

In this section, we will give a high-level summary of our model and inference method. For a more detailed description of the theory underpinning the neural likelihood-ratio method, we refer readers to \cite{cranmer2015approximating}, while for more details on its application to analyzing subhalo density slopes in strong lensing images, we refer readers to \cite{zhang2022}. 

\subsection{Inference method}
\label{subsec:infmethod}

Suppose $\theta$ denotes the parameters of our interest and $x$ denotes the observed data. The core idea of likelihood-ratio estimation is training a classifier to distinguish between samples from two different probability distributions: the joint data-parameter distribution $p(x, \theta)$, which is the distribution of our interest, and the product of the marginal distributions of the data and the parameter $p(x) p(\theta)$. In our case, $x$ corresponds to observed strong lensing images, while $\theta$ is the subhalo density slope $\gamma$ underlying each image. We train a neural network as a classifier to learn the decision function \add{$s(x, \theta) = p(x,  \theta)/\left(p(x, \theta) + p(x) p(\theta)\right)$, which is in one-to-one correspondence with the likelihood ratio $r(x\mid\theta) = p(x,  \theta)/(p(x) p(\theta))$} as follows:  
\begin{align}
     r(x\mid\theta) = \frac{s(x, \theta)}{1 - s(x, \theta)}.
     \label{eq:llr}
\end{align}
This allows us to convert a likelihood inference task to a classification task~\citep{cranmer2015approximating, baldi2016parameterized, mohamed2016learning}. At test time, to compute the likelihood-ratio profile as a function of $\gamma$ for a given lensed image, we obtain the classifier logits for a linearly-spaced array of input $\gamma$ values. The likelihood-ratio estimation method is amortized: after spending an initial overhead for model training, minimal computational cost is needed during inference.

If we have an ensemble of strong lensing observations $\{x\}$ that are independently and identically distributed when conditioned on $\gamma$, then we can obtain their combined likelihood ratio by computing the product of the individual likelihood ratios, 
\begin{align}
    \hat{r} (\{x\} \mid \gamma) = \prod_{i} \hat{r}(x_i \mid \gamma). 
    \label{eq:combine_r}
\end{align}
This offers a way for us to efficiently combine results of multiple observations with little additional computational cost. 

\subsection{Uncertainty quantification}
\label{subsec:uncertainty}

If our likelihood-ratio estimator is a perfect classifier, then the test-statistic $2 \left(\ln  \hat{r}_\mathrm{MLE} - \ln \hat{r}\right)$ should be $\chi^2$-distributed~\citep{Wilks:1938dza}, where $\ln \hat{r}$ is the log-likelihood evaluated at the true $\gamma$ and $\ln  \hat{r}_\mathrm{MLE}$ is the log-likelihood at the maximum likelihood estimate (MLE) $\gamma_{\mathrm{MLE}}$. However, we found that with our imperfect classifier, the test-statistic distribution deviates slightly from a true $\chi^2$. Therefore, instead of quoting the $68\%$ uncertainty interval using a $\chi^2$ distribution, we empirically determine the threshold for the 68\% confidence interval (CI) of the test-statistic. In practical terms, we do this by computing the test-statistics of many samples and then determining a threshold value under which approximately 68\% of the test-statistics of these samples are included. Then, for a likelihood-ratio profile, the $\gamma$ values whose likelihood-ratios evaluate to this threshold determine the upper and lower uncertainties on the MLE. Because we found that combining different numbers of likelihood-ratios leads to slightly different test-statistic distributions, this empirical threshold is determined separately for combining different numbers of images. This uncertainty quantification procedure ensures that approximately 68\% of the ground truths fall within the uncertainties quoted, and is used to determine the error bars presented in Sec.~\ref{sec:results}. 

\subsection{Model and training details}
\label{subsec:train_details}

For our application, we use as our classifier a modified version of a common computer vision model, the ResNet-50 convolutional neural network implemented in \texttt{PyTorch}~\citep{he2016deepresiduallearning, NEURIPS2019_9015}. We add a sigmoid projection after all of the dense layers in the ResNet in order to output the classification score $\hat s(x,\theta)$. At training time, we append the true $\gamma$ for each input image to the latent vector after the convolutional layers in order to ensure that the model learns the true label. At test time, we instead append test $\gamma$ values in order to obtain likelihood-ratio estimates over a range of $\gamma$. Our training objective is the canonical binary cross-entropy loss for classification. \add{We provide more details of our customized ResNet-50 architecture in Appendix~\ref{appx:a}.}

To help with model convergence, we pre-process our training set images. We normalize image pixel values to having zero mean and unit standard deviation across the training set; in addition, we normalize the $\gamma$ values to zero mean. To ensure consistency at test time, we use the training set mean and standard deviation to whiten our test data. 

We use the AdamW optimizer~\citep{kingma2014adam, loshchilove2017adamw} with an initial learning rate of $10^{-3}$. We follow a learning rate schedule that decays by an order of magnitude when the validation loss stagnates for 3 epochs, followed by a 2-epoch cool-down period. We use a batch size of 1000 based on the maximum GPU memory available. There are 5,000,000 mock images in our training set and 1,000 in our validation set, all of which are generated using the forward model described in Sec.~\ref{sec:data}. Training terminates when the validation loss plateaus under a threshold of $10^{-3}$. We carried out our neural network training on NVIDIA V100 GPUs for $\sim$20 epochs, with each epoch taking $\sim$5 hours. We found that scaling up the size of the training set and the model complexity significantly improved the model performance during inference, and we expect there to be more improvement if the computing resource availability allows for more up-scaling. 

\section{Results}
\label{sec:results}

\begin{figure}
    \centering
    \includegraphics[width=\linewidth]{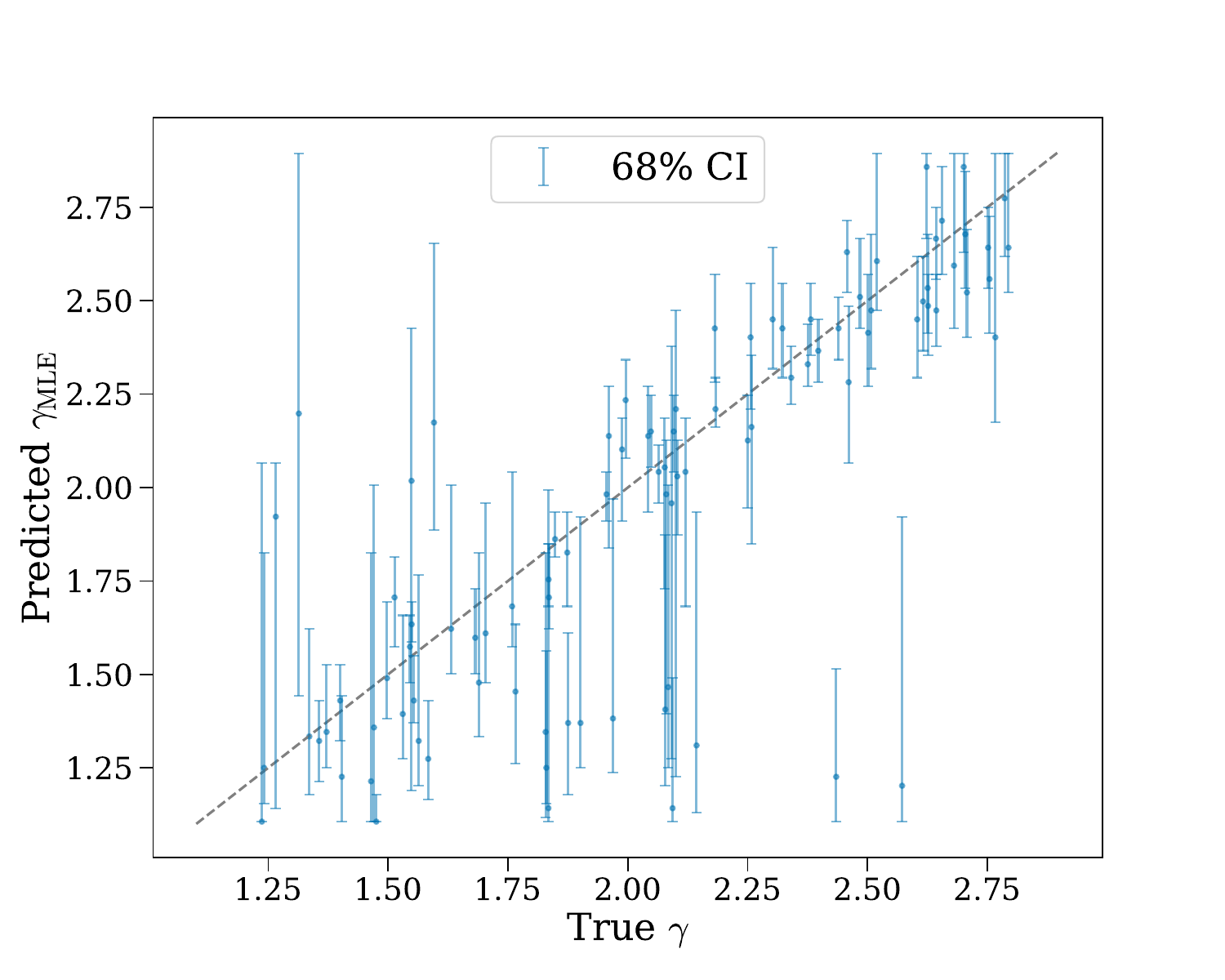}
    \caption{Scatter plot of the maximum-likelihood estimates $\gamma_{\mathrm{MLE}}$ and their associated 68\% confidence intervals (as discussed in Sec.~\ref{subsec:uncertainty}) predicted using the trained likelihood-ratio estimator compared to the true underlying $\gamma$ of 93 test images. The images contain EPL subhalos with $M_{200} \in [10^7, 10^{10}]$ $\mathrm M_{\odot}$ and $N_{\mathrm{sub}} \sim \mathcal U\{0, 1800\}$. The model was trained on images containing EPL subhalos with $M_{200} \in [10^7, 10^{10}]$ $\mathrm M_{\odot}$ and $N_{\mathrm{sub}} \sim \mathcal U\{0, 3000\}$.}
    \label{fig:scatter_validation}
\end{figure}

After our model has been trained, we first need to evaluate its convergence. To do this, we compare model predictions of the subhalo density slope with their ground truths using individual images in our validation set. In Fig.~\ref{fig:scatter_validation}, we show the maximum likelihood estimate (along with their 68\% uncertainties) compared to the true $\gamma$ for 93 images with $1.2 < \gamma < 2.8$. 

The validation images have parameters drawn from the same distributions as the training set, except for $N_{\mathrm{sub}}$, which is drawn from $\mathcal{U}\{0, 1800\}$ to simulate a more realistic substructure fraction. \add{Note that because each image in our training and validation set is labeled with a true underlying $\gamma$ value for EPL subhalos, we ideally would like the neural network to predict the ground truths and be agnostic to the number of subhalos. Therefore, having a more realistic number of subhalos in our validation set serves as a way for us to ensure that changing the number of subhalos does not incur a bias in our neural network predictions.} The source galaxy images used in validation were held out in training. These images contain EPL subhalos whose true underlying subhalo density slopes are known in the forward simulation pipeline, making this direct comparison possible. As shown in Fig.~\ref{fig:scatter_validation}, our model predictions follow the ground truths in trend. This demonstrates that despite the addition of several layers of complexities into training images, our neural network remains sensitive to the signature imprinted on strong lensing images by changes in the subhalo density slope. However, the relatively large confidence intervals indicate that the constraining power of individual images is limited, which makes it imperative to combine multiple images for inference. Note that the uncertainties are larger at the lower end of the $\gamma$ range because smaller $\gamma$ values indicate less concentrated subhalos, which leave less detectable signatures in the lensed images.

\begin{figure}
    \centering
    \includegraphics[width=\linewidth]{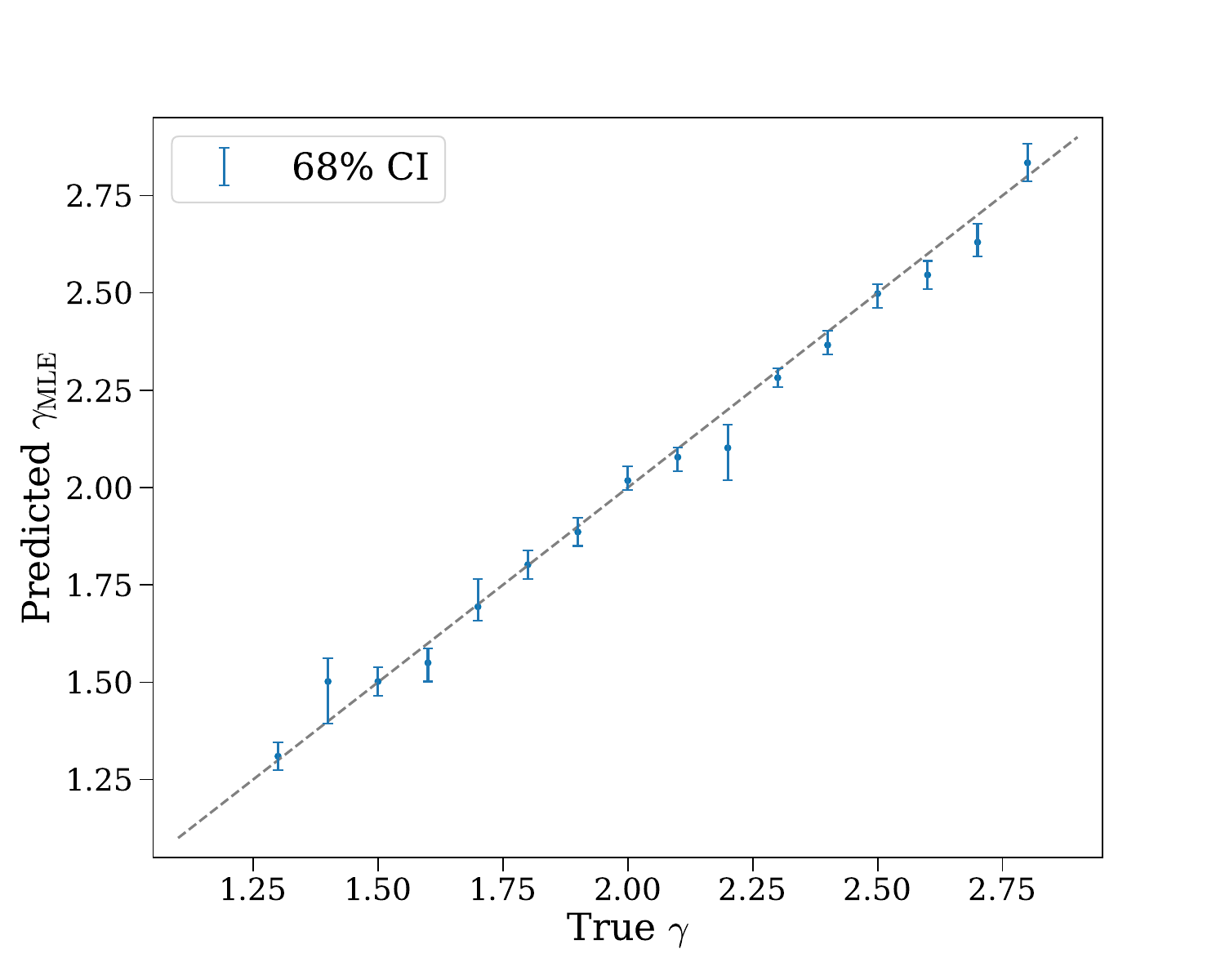}
    \caption{Maximum-likelihood estimated $\gamma_{\mathrm{MLE}}$ and associated 68\% confidence intervals predicted from combined likelihoods of sets of 13 images containing EPL subhalos compared to the true $\gamma$ underlying each set of images (with $M_{200} \in [10^7, 10^{10}]$ $\mathrm M_{\odot}$ and $N_{\mathrm{sub}} \sim \mathcal U\{0, 1800\}$). The model was trained on images containing EPL subhalos with $M_{200} \in [10^7, 10^{10}]$ $\mathrm M_{\odot}$ and $N_{\mathrm{sub}} \sim \mathcal U\{0, 3000\}$.}
    \label{fig:combine_epl}
\end{figure}

In addition, we check the model predictions of combining likelihood ratios of multiple images. In Fig.~\ref{fig:combine_epl}, we show $\gamma_{\mathrm{MLE}}$ compared to the ground truth $\gamma$ for combined likelihoods of sets of 13 images, with each set sharing the same underlying slope $\gamma$. Note that we still simulate the natural spread in $\gamma_i$, so the slope for each subhalo varies slightly. These images share the same parameter distributions as the images used in Fig.~\ref{fig:scatter_validation} except that the source galaxies come from the held-out set for validation. In Fig.~\ref{fig:combine_epl}, we see that the MLE predictions of our model closely follow the ground truths with relatively small error bars. Comparing the uncertainties between Fig.~\ref{fig:scatter_validation} and Fig.~\ref{fig:combine_epl}, we see that combining images significantly improves constraining power. 

\subsection{Simulated images with NFW subhalos}
\label{subsec:nfw_sims}

To obtain the expected subhalo density slopes under the cold dark matter model, we simulate images containing subhalos and LoS halos following the Navarro–Frenk–White (NFW) profile~\citep{navarro1997nfw}. Its radial density profile given by: 
\begin{align}
    \rho(r) = \frac{\rho_0}{\frac{r}{r_s}\left(1 + \frac{r}{r_s}\right)^2},
\end{align}
where $r$ is the distance from the center of a subhalo, and the normalization $\rho_0$ and the scale radius $r_s$ are free parameters. The NFW profile is the most common fit for the universal density profile of halos from CDM N-body simulations. In addition to a normalization and a scale radius, the NFW profile can also be parametrized by the (sub)halo mass $M_{200}$ and concentration $c_{200}$. The concentration $c_{200}$ relates to the scale radius and virial radius $r_{200}$ (as defined in Sec.~\ref{subsubsec:subhalo_and_los}) following $r_{200} = c_{200} r_s$. In our simulated images with NFW subhalos, we relate $M_{200}$ and $c_{200}$ with a mass-concentration relation extrapolated from \citet{dutton2014cdmhalo}, which is an empirical relation determined using halos in CDM simulations. We add a dex scatter of $\mathcal{N}(0, 0.1)$ to the mass-concentration relation for each subhalo in order to mimic the natural spread in the relationship. Note that a dex scatter of 0.1 corresponds to a $\sim 26\%$ variation in concentration. If we denote the CDM mass-concentration as $f_{\mathrm{CDM}}(M_{200})$, then we can modify the CDM mass-concentration relation by multiplying it by a constant factor (which will be referred to as concentration multiplicative factor) in order to simulate different density slopes of subhalo populations. \add{In other words, this means that for a given concentration multiplicative factor $a$, we set the concentration of a subhalo with mass $M_{200}$ to be $a \times f_{\mathrm{CDM}}(M_{200})$.} To test the robustness of our neural network with as realistic images as possible, we add subhalos everywhere in the image in these test sets. 

In addition, due to tidal stripping from the host halo, subhalos typically lose mass in their outer region~\citep{hayashi2003, diemand2011}. This means that, instead of an NFW profile, they can be more realistically modeled by a truncated NFW (tNFW) profile. The tNFW profile is parametrized by the NFW parameters as well as a truncation radius $r_{t}$: 
\begin{align}
    \rho(r) = \frac{r_t^2 + r^2}{r_t^2} \frac{\rho_0}{r/{r_s}\left(1 + r/{r_s}\right)^2}. 
\end{align}
Because the truncation steepens the subhalo density profile past the truncation radius, we expect tNFW subhalos to have steeper power-law density slopes than their NFW counterparts. To model the tNFW subhalos in our pipeline, we first determine their NFW parameters following the procedure described for NFW subhalos and subsequently set their truncation radii. The choice of truncation radii affects subhalo density profiles in the intermediate and outer region and thereby their measured slopes~\citep{sengul2022probing}. For our test images, we set the truncation radii following~\cite{wagner2022images}:
\begin{align}
    r_t = 1.4\left(\frac{M_{200}}{m_{\mathrm{trunc, pivot}}}\right)^{\frac{1}{3}}\left(\frac{r_{\mathrm{sub}}}{r_{\mathrm{trunc, pivot}}}\right)^{\frac{2}{3}},
\end{align}
with $m_{\mathrm{trunc, pivot}} = 10^7\mathrm{M_{\odot}}$ and $r_{\mathrm{trunc, pivot}} = 50$~kpc. Here, $M_{200}$ is the subhalo mass and $r_{\mathrm{sub}}$ is the distance of the subhalo from the main halo center. Note that in the test sets where subhalos are modeled with tNFW, LoS halos are still modeled with the NFW profile as they experience less tidal stripping than subhalos. 

One question that might arise is why our trained likelihood-ratio estimator can be applied to images with (t)NFW subhalos and LoS halos even though the training set only contains their EPL counterparts. The justification for this has been demonstrated in \cite{sengul2022probing} and \cite{zhang2022}: given limited resolution and appropriate noise level, the observable changes in the surface brightness due to the presence of (t)NFW subhalos can be well approximated by that of a power-law profile subhalo.

Because the density slopes of (t)NFW subhalos vary with mass (\textit{i.e.} larger masses have more extended density profiles and thereby smaller density slopes), the intrinsic stochasticity in the masses of a (t)NFW subhalo population introduces intrinsic aleatoric uncertainty into the slope measurement. To account for this uncertainty in each of our measurements, we generate 100 separate sets of images with shared properties and then obtain the MLE of each combined likelihood; using the set of 100 MLEs, we empirically determine the 68\% confidence intervals.

\begin{figure}
    \centering
    \includegraphics[width=\linewidth]{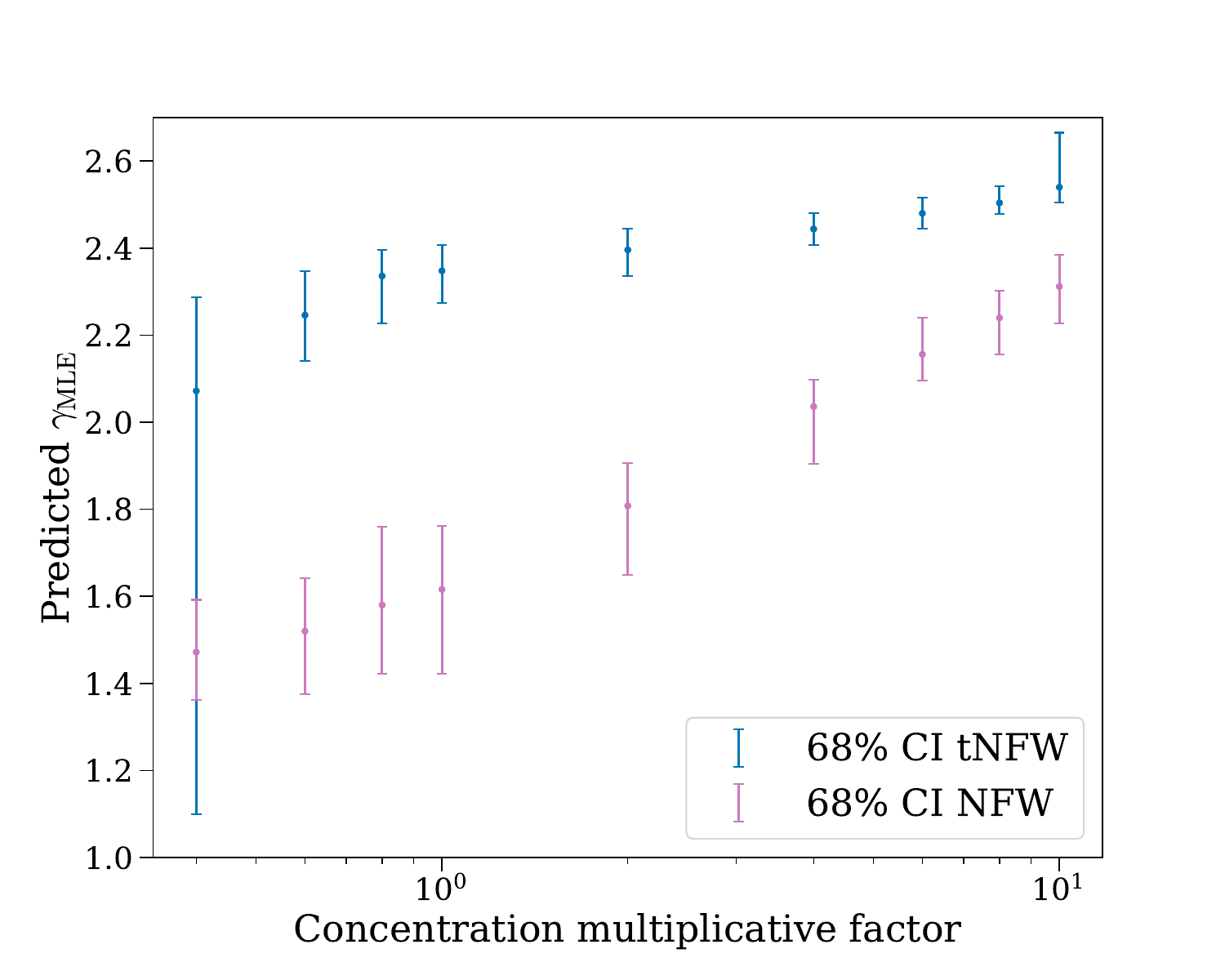}
    \caption{The median maximum-likelihood predictions (scatter points) and $68\%$ confidence intervals (error bars) obtained by combining 13 images of $M_{200}  \in [10^7, 10^{10}]\mathrm M_{\odot}$ NFW subhalos (magenta) or tNFW subhalos (blue) as a function of the concentration multiplicative factor.}
    \label{fig:combine_nfw}
\end{figure}

In Fig.~\ref{fig:combine_nfw}, we sample an array of varying concentration multiplicative factors and show our model MLE from the combined likelihood of 13 images that contain (t)NFW subhalos and LoS halos at each multiplicative factor. As expected, subhalos with higher concentrations lead to larger $\gamma$ predictions. In particular, the data points for a concentration multiplicative factor of 1 correspond to the expected subhalo density slope measurements under the CDM model, and they will be compared with the predicted slope of the HST observations in Sec.~\ref{subsec:compare_hst}. From the figure, we find that tNFW subhalos in general produce higher density slope measurements than NFW subhalos, consistent with findings presented in~\cite{sengul2022probing}. 

\subsection{Result with HST images}
\label{subsec:compare_hst}

\begin{figure}
    \centering
    \includegraphics[width=\linewidth]{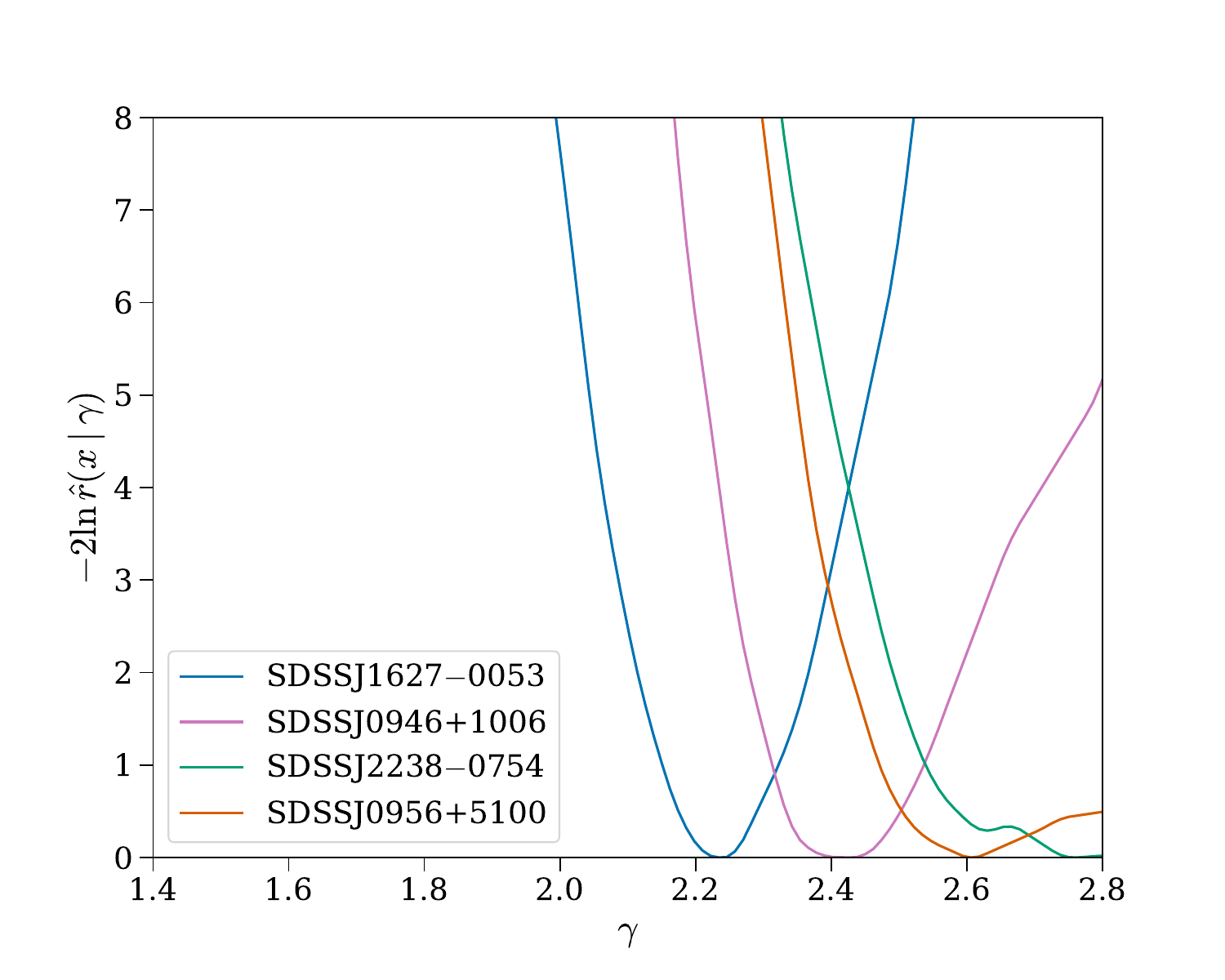}
    \caption{Representative examples of the individual likelihood-ratio test statistic profiles for HST images. Each profile is labeled with the name of its corresponding lens system.} 
    \label{fig:hst_indiv}
\end{figure}

Having done model validation and obtained the expected density slope of CDM subhalos, we will now use our model to infer the subhalo slope of observed HST strong lensing images, which are described in Sec.~\ref{subsec:hst_data}. These images are masked at the edges and whitened with the training mean and standard deviation before being fed into our neural network. In Fig.~\ref{fig:hst_indiv}, we show the individual likelihood-ratio test statistic profiles for several of the HST observations. In comparison with the predictions of NFW and tNFW subhalos under the CDM model, as discussed in Sec.~\ref{subsec:nfw_sims}, we see that the predicted slopes of these HST observations are larger than the predicted slopes of the CDM model. 

\begin{figure}
    \centering
    \includegraphics[width=\linewidth]{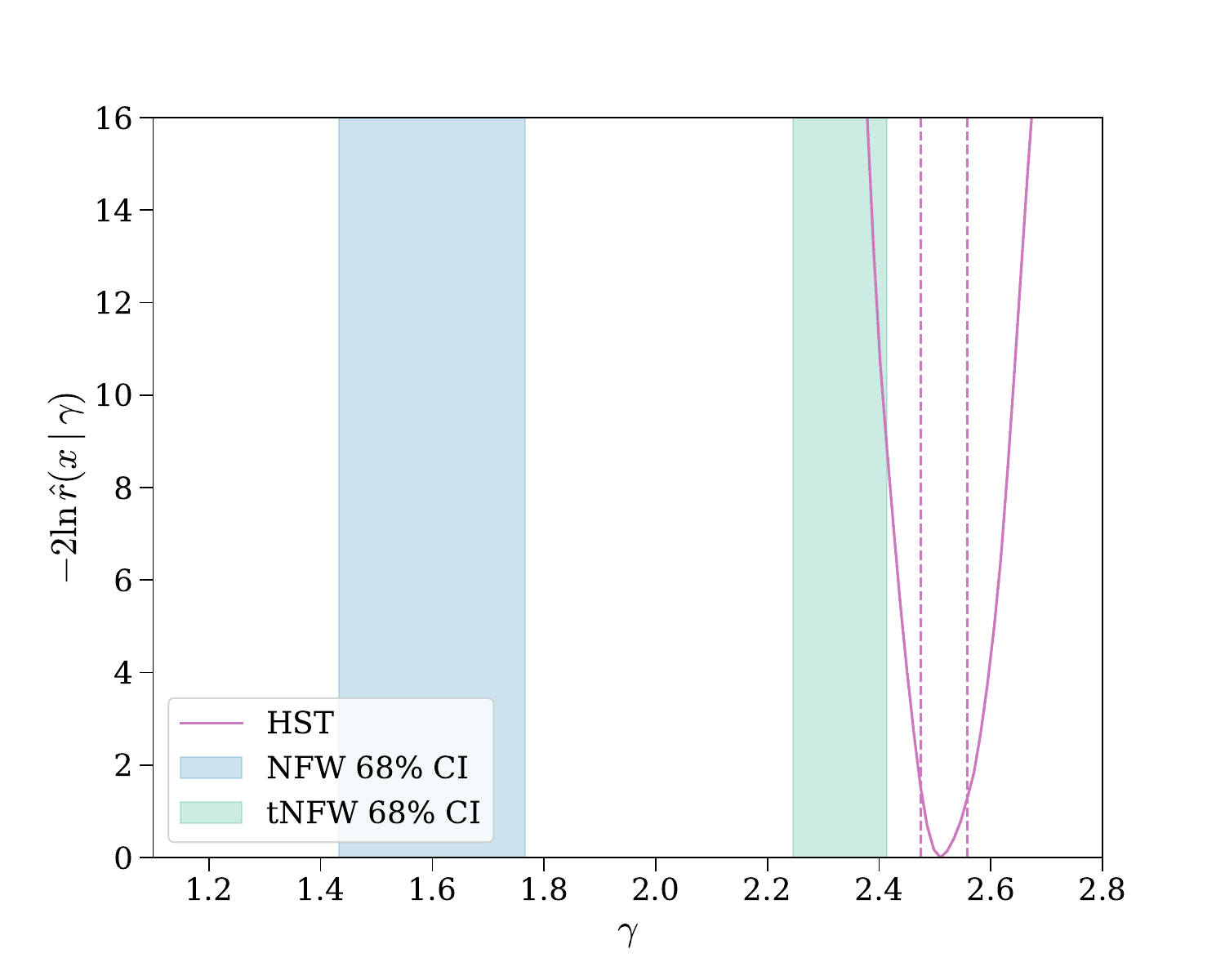}
    \caption{The 68\% confidence interval of combined likelihood-ratio test statistic profiles of 13 images containing $M_{200} \in [10^7, 10^{10}]$ $\mathrm M_{\odot}$ NFW subhalos (blue) and tNFW subhalos (green), both with a concentration multiplicative factor of 1, as well as the combined likelihood-ratio test statistic profile of the 13 HST images shown in Fig.~\ref{fig:hst_images} (magenta). The uncertainties corresponding to the 68\% confidence interval are shown in dashed lines for the likelihood-ratio test statistic profile.}
    \label{fig:combine_hst}
\end{figure}

We subsequently combine these individual likelihood ratios using Eq.~\ref{eq:combine_r} in order to obtain tighter constraints. In Fig.~\ref{fig:combine_hst}, we show the combined likelihood-ratio test statistic profile for all of the 13 images. From this profile, we get a measurement of the subhalo density slope of $\gamma_{\mathrm{MLE}} = 2.51 ^{-0.04}_{+0.05}$, with the quoted uncertainties indicating the 68\% credible interval shown in dotted lines. 

In the same figure, we also show the 68\% confidence intervals for the combined likelihood-ratio test statistic profiles of 13 images containing NFW subhalos or tNFW subhalos. These correspond to data points shown in Fig.~\ref{fig:combine_nfw} for a concentration multiplicative factor of 1. Comparing the slope prediction of the HST images with that of the simulated images with NFW subhalos, we see that the measured density slope of the HST data is significantly steeper than the expected slope under the assumption that CDM subhalos follow an NFW profile. The predicted slope of the images containing tNFW subhalos is also less than the HST measurement, but their difference is less statistically significant than that with the NFW prediction. While surprising, this is in agreement with previous works that also measured a higher than expected concentration~\citep{minor2021unexpected, sengul2022probing}. In particular, our 13 HST images include the SDSSJ0946+1006 system analyzed by~\cite{minor2021unexpected}, which measured a much higher concentration than the CDM prediction. The individual likelihood-ratio test statistic profile for the same system in our analysis is shown in Fig.~\ref{fig:hst_indiv}, and it is in broad agreement with the result in their work. It is also worth noting that our method provides a stronger constraint due to the neural network's ability to efficiently combine multiple observations. \add{It would also be useful to compare our results with those obtained by \cite{sengul2022probing} of the JVAS B1938+666 lens system, but to our knowledge, there is no suitable HST observation of this lens system that matches our training set configuration. Thus, we leave this for future work when more observations become available.}

One possible explanation of the difference between our result and the CDM predictions lies in the assumptions made in our subhalo modeling. Several assumptions about subhalo density profiles went into modeling the lens system in the image; in particular, the density profile parametrization and the choice of mass-concentration relation affect the predicted slope measurements of subhalos under the CDM model. Modeling these properties for subhalos is an onging area of research~\citep{green2021}, and an improved understanding of subhalo profiles may change the predicted CDM density slopes. Another possible reconciliation is accounting for the selection effects. Subhalos with steeper density slopes are more concentrated and, therefore, are easier to detect in observations. Within our current resolution constraint and noise level, the less concentrated smaller subhalos are not detectable, hence biasing our statistics. This effect of the selection function on slope measurements is important, and we leave a careful study of it for future work, when more observations become available from ongoing and upcoming surveys. 

\section{Conclusions and outlook}
\label{sec:conclusion}

Observations at sub-galactic scale are essential for examining alternate dark matter models and contrasting them against the standard CDM model. Among the small-scale observables, subhalos provide a promising avenue for dark matter studies. In addition to constraining the subhalo mass function, studying the subhalo density slope (concentration) can help to potentially differentiate various classes of dark matter models. Subhalo properties can be probed by analyzing strong gravitational lensing images. Traditional strong lensing image analyses model individual subhalos through a forward modeling pipeline, but this process can only provide limited statistics; to model more subhalos in a system or to combine statistics from many images, direct lens modeling becomes computationally infeasible. 

The rapid progress in machine learning enables the development of techniques that have the power to leverage the collective effect of subhalo populations in strong lensing images, as well as to efficiently analyze a large ensemble of observations. Despite showcases of success on simulated images, many of these machine learning methods require further validation and improvements before they can be successfully applied to real strong lensing observations. 

In this work, we built upon the likelihood-ratio estimation method developed in~\cite{zhang2022} and trained a neural network capable of making inference from observed strong lensing images. To make the leap from mock to real images, we added numerous layers of realism in the forward pipeline of the training set. This includes complexifying the lens model to account for the lens light, multipole moments as well as external shear, incorporating realistic noise levels, and adding line-of-sight halos. We demonstrated that the likelihood-ratio estimator retains its sensitivity to changes in the subhalo density slope in simulated strong lensing, even after adding these layers of realism. Furthermore, we obtained the expected subhalo density slope measurements in simulations under the CDM model. This measurement comes from using our trained neural network to predict the slope of simulated lensing images containing (t)NFW subhalos that follow a mass-concentration relation derived from CDM simulations. Finally, we measured the subhalo slopes of a set of 13 HST observations and statistically combined their constraints. By comparing the subhalo slope in the HST observations with the measurement from simulated CDM images, we found an unexpectedly high slope measurement in the HST observations, in tension with CDM predictions.  

\add{Several recent works in cluster lensing have also suggested that substructures in galaxy clusters are more compact than expected of the CDM model~\citep{meneghetti2020, meneghetti2022,meneghetti2023}.} \add{Combined with several similar results in the literature, our} measurement has important implication for dark matter studies as it may motivate more careful examination of alternate dark matter models. The most common alternatives to CDM, the warm dark matter model and many self-interacting dark matter models, predict a lower than CDM subhalo density slope and would exacerbate the tension that we observe~\citep{lovell2012wdm, lovell2014wdm, vogelsberger2012sidm, rocha2013sidm, kahlhoefer2019sidm}. However, certain self-interacting dark matter models (\textit{e.g.} with large self-interacting cross sections~\citep{nishikawa2020}) also predict that SIDM subhalos can undergo core collapses that result in unusually concentrated inner profiles in a time-scale relevant for observations today~\citep{lyndenbell1968, 2000kochanek, 2002colin, elbert2015core, nadler2023}. \add{This gravitothermal core collapse due to dark matter self interactions has been suggested as a possible explanation of these high density central regions in cluster galaxies~\citep{core_collapse}.} \add{ Resolving galactic subhalos in simulations is harder due to their lower masses. A hybrid approach in \cite{core_collapse_hybrid}, which includes a combination of semi-analytical methods and N-body simulations has shown that some SIDM models can produce subhalos with collapsed cores at subgalactic mass scales ($<10^{10}\, \mathrm{M}_\odot$).} This phenomenon provides a possible explanation for the high subhalo density slope that we measured. Based on our work, it is still not possible to pinpoint the mechanism that causes this outlier measurement from the CDM model, but there are several directions of future work that can take us closer to answering this question. For instance, one can study the subhalo slope predictions under different microphysical dark matter models and compare them with the predictions from observed lensing images. In addition, one can examine the effect of assumptions about CDM subhalo properties on the likelihood-ratio estimator's slope predictions. As more lensing systems are expected to be discovered with upcoming surveys (and followed up by observations), the likelihood-ratio estimator will be a valuable tool for obtaining more measurements to help elucidate the nature of dark matter. 

\section*{Acknowledgements}
We thank Simon Birrer and Siddharth Mishra-Sharma for helpful discussions.
This work is supported by the National Science Foundation under Cooperative Agreement PHY-2019786 (The NSF AI Institute for Artificial Intelligence and Fundamental Interactions, \url{http://iaifi.org/}).
The computations in this paper were run on the FASRC Cannon cluster supported by the FAS Division of Science Research Computing Group at Harvard University.

\section*{Data Availability}

The 13 HST images analyzed in this work can be downloaded from \url{https://mast.stsci.edu/portal/Mashup/Clients/Mast/Portal.html}. The code used to produce the results shown in this paper is available at \url{https://github.com/gemyxzhang/neural-subhalo-slope-data}.

\appendix

\add{\section{Model architecture}
\label{appx:a}
 We describe in this appendix the customized ResNet-50 architecture used in this work. The original ResNet-50 model used in computer vision consists of a series of convolution blocks followed by pooling and dense layers. We made two modifications to this model for our inference task. Firstly, we append the truth label $\gamma$ of each image during training to the flattened latent space vector after the convolution blocks, as indicated by the top arrow in Fig.~\ref{fig:model}. This ensures that the neural network incorporates information about $\gamma$ into its prediction. In addition, we add a logistic activation function after the last layer of ResNet-50 to ensure that the final output is a valid classification score $\hat{s}(\gamma, x)$ (\textit{i.e.} between 0 and 1). As discussed in Sec.~\ref{subsec:train_details}, when we train the neural network as a classifier, the value given by the ResNet before the logistic activation gives us the log likelihood estimate $\ln{\hat{r}}$, as indicated in Fig.~\ref{fig:model}. 
\begin{figure}
    \centering
    \includegraphics[width=\linewidth]{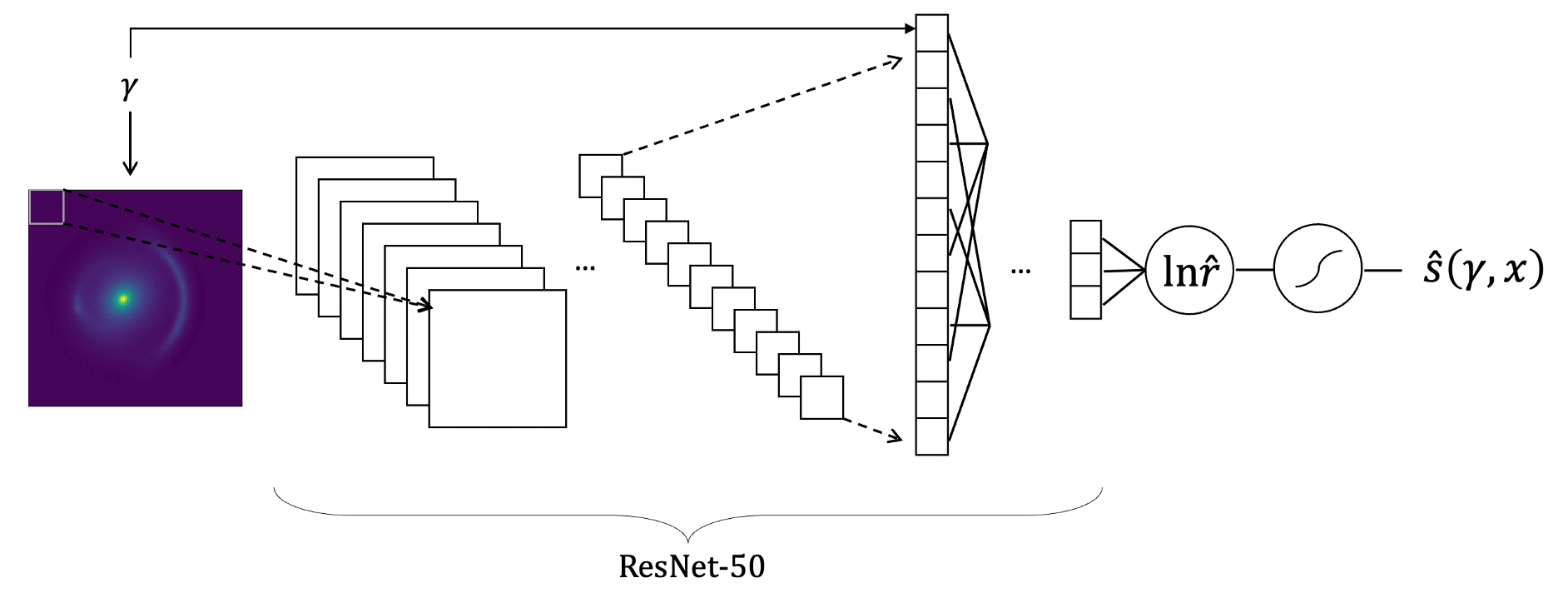}
    \caption{Graphical illustration of the neural network architecture used in this work.}
    \label{fig:model}
\end{figure}}
\label{lastpage}
\clearpage
\bibliography{references} 

\begin{thebibliography}{}
\makeatletter
\relax
\def\mn@urlcharsother{\let\do\@makeother \do\$\do\&\do\#\do\^\do\_\do\%\do\~}
\def\mn@doi{\begingroup\mn@urlcharsother \@ifnextchar [ {\mn@doi@} {\mn@doi@[]}}
\def\mn@doi@[#1]#2{\def\@tempa{#1}\ifx\@tempa\@empty \href {http://dx.doi.org/#2} {doi:#2}\else \href {http://dx.doi.org/#2} {#1}\fi \endgroup}
\def\mn@eprint#1#2{\mn@eprint@#1:#2::\@nil}
\def\mn@eprint@arXiv#1{\href {http://arxiv.org/abs/#1} {{\tt arXiv:#1}}}
\def\mn@eprint@dblp#1{\href {http://dblp.uni-trier.de/rec/bibtex/#1.xml} {dblp:#1}}
\def\mn@eprint@#1:#2:#3:#4\@nil{\def\@tempa {#1}\def\@tempb {#2}\def\@tempc {#3}\ifx \@tempc \@empty \let \@tempc \@tempb \let \@tempb \@tempa \fi \ifx \@tempb \@empty \def\@tempb {arXiv}\fi \@ifundefined {mn@eprint@\@tempb}{\@tempb:\@tempc}{\expandafter \expandafter \csname mn@eprint@\@tempb\endcsname \expandafter{\@tempc}}}

\bibitem[\protect\citeauthoryear{{Amorisco} et~al.,}{{Amorisco} et~al.}{2022}]{amorisco2022}
{Amorisco} N.~C.,  et~al., 2022, \mn@doi [\mnras] {10.1093/mnras/stab3527}, \href {https://ui.adsabs.harvard.edu/abs/2022MNRAS.510.2464A} {510, 2464}

\bibitem[\protect\citeauthoryear{{Anau Montel}, {Coogan}, {Correa}, {Karchev}  \& {Weniger}}{{Anau Montel} et~al.}{2022}]{2022arXiv220509126A}
{Anau Montel} N.,  {Coogan} A.,  {Correa} C.,  {Karchev} K.,   {Weniger} C.,  2022, arXiv e-prints, \href {https://ui.adsabs.harvard.edu/abs/2022arXiv220509126A} {p. arXiv:2205.09126}

\bibitem[\protect\citeauthoryear{{Anderson} \& {King}}{{Anderson} \& {King}}{2000}]{anderson2000}
{Anderson} J.,  {King} I.~R.,  2000, \mn@doi [\pasp] {10.1086/316632}, \href {https://ui.adsabs.harvard.edu/abs/2000PASP..112.1360A} {112, 1360}

\bibitem[\protect\citeauthoryear{{Auger}, {Treu}, {Bolton}, {Gavazzi}, {Koopmans}, {Marshall}, {Bundy}  \& {Moustakas}}{{Auger} et~al.}{2009}]{auger2009slacs}
{Auger} M.~W.,  {Treu} T.,  {Bolton} A.~S.,  {Gavazzi} R.,  {Koopmans} L.~V.~E.,  {Marshall} P.~J.,  {Bundy} K.,   {Moustakas} L.~A.,  2009, \mn@doi [\apj] {10.1088/0004-637X/705/2/1099}, \href {https://ui.adsabs.harvard.edu/abs/2009ApJ...705.1099A} {705, 1099}

\bibitem[\protect\citeauthoryear{{Baldi}, {Cranmer}, {Faucett}, {Sadowski}  \& {Whiteson}}{{Baldi} et~al.}{2016}]{baldi2016parameterized}
{Baldi} P.,  {Cranmer} K.,  {Faucett} T.,  {Sadowski} P.,   {Whiteson} D.,  2016, \mn@doi [European Physical Journal C] {10.1140/epjc/s10052-016-4099-4}, \href {https://ui.adsabs.harvard.edu/abs/2016EPJC...76..235B} {76, 235}

\bibitem[\protect\citeauthoryear{{Barkana}}{{Barkana}}{1998}]{barkana1998epl}
{Barkana} R.,  1998, \mn@doi [\apj] {10.1086/305950}, \href {https://ui.adsabs.harvard.edu/abs/1998ApJ...502..531B} {502, 531}

\bibitem[\protect\citeauthoryear{{Bechtol} et~al.,}{{Bechtol} et~al.}{2019}]{bechtol2019lsst}
{Bechtol} K.,  et~al., 2019, \baas, \href {https://ui.adsabs.harvard.edu/abs/2019BAAS...51c.207B} {51, 207}

\bibitem[\protect\citeauthoryear{{Birrer} \& {Amara}}{{Birrer} \& {Amara}}{2018}]{birrer2018lenstronomy}
{Birrer} S.,  {Amara} A.,  2018, \mn@doi [Physics of the Dark Universe] {10.1016/j.dark.2018.11.002}, \href {https://ui.adsabs.harvard.edu/abs/2018PDU....22..189B} {22, 189}

\bibitem[\protect\citeauthoryear{{Birrer}, {Amara}  \& {Refregier}}{{Birrer} et~al.}{2015}]{birrer2015gravitational}
{Birrer} S.,  {Amara} A.,   {Refregier} A.,  2015, \mn@doi [\apj] {10.1088/0004-637X/813/2/102}, \href {https://ui.adsabs.harvard.edu/abs/2015ApJ...813..102B} {813, 102}

\bibitem[\protect\citeauthoryear{Birrer, Amara  \& Refregier}{Birrer et~al.}{2017}]{birrer2017lensingsubs}
Birrer S.,  Amara A.,   Refregier A.,  2017, \mn@doi [JCAP] {10.1088/1475-7516/2017/05/037}, 05, 037

\bibitem[\protect\citeauthoryear{{Bode}, {Ostriker}  \& {Turok}}{{Bode} et~al.}{2001}]{bode2001wdm}
{Bode} P.,  {Ostriker} J.~P.,   {Turok} N.,  2001, \mn@doi [\apj] {10.1086/321541}, \href {https://ui.adsabs.harvard.edu/abs/2001ApJ...556...93B} {556, 93}

\bibitem[\protect\citeauthoryear{{Bolton}, {Burles}, {Koopmans}, {Treu}, {Gavazzi}, {Moustakas}, {Wayth}  \& {Schlegel}}{{Bolton} et~al.}{2008}]{bolton2008slacs}
{Bolton} A.~S.,  {Burles} S.,  {Koopmans} L. V.~E.,  {Treu} T.,  {Gavazzi} R.,  {Moustakas} L.~A.,  {Wayth} R.,   {Schlegel} D.~J.,  2008, \mn@doi [\apj] {10.1086/589327}, \href {https://ui.adsabs.harvard.edu/abs/2008ApJ...682..964B} {682, 964}

\bibitem[\protect\citeauthoryear{{Brehmer}, {Mishra-Sharma}, {Hermans}, {Louppe}  \& {Cranmer}}{{Brehmer} et~al.}{2019}]{brehmer2019mining}
{Brehmer} J.,  {Mishra-Sharma} S.,  {Hermans} J.,  {Louppe} G.,   {Cranmer} K.,  2019, \mn@doi [\apj] {10.3847/1538-4357/ab4c41}, \href {https://ui.adsabs.harvard.edu/abs/2019ApJ...886...49B} {886, 49}

\bibitem[\protect\citeauthoryear{Brennan, Benson, Cyr-Racine, Keeton, Moustakas  \& Pullen}{Brennan et~al.}{2019}]{brennan2018quantifying}
Brennan S.,  Benson A.~J.,  Cyr-Racine F.-Y.,  Keeton C.~R.,  Moustakas L.~A.,   Pullen A.~R.,  2019, \mn@doi [Mon. Not. Roy. Astron. Soc.] {10.1093/mnras/stz1607}, 488, 5085

\bibitem[\protect\citeauthoryear{{Brewer}, {Huijser}  \& {Lewis}}{{Brewer} et~al.}{2016}]{brewer2016transdim}
{Brewer} B.~J.,  {Huijser} D.,   {Lewis} G.~F.,  2016, \mn@doi [\mnras] {10.1093/mnras/stv2370}, \href {https://ui.adsabs.harvard.edu/abs/2016MNRAS.455.1819B} {455, 1819}

\bibitem[\protect\citeauthoryear{{Col{\'\i}n}, {Avila-Reese}, {Valenzuela}  \& {Firmani}}{{Col{\'\i}n} et~al.}{2002}]{2002colin}
{Col{\'\i}n} P.,  {Avila-Reese} V.,  {Valenzuela} O.,   {Firmani} C.,  2002, \mn@doi [\apj] {10.1086/344259}, \href {https://ui.adsabs.harvard.edu/abs/2002ApJ...581..777C} {581, 777}

\bibitem[\protect\citeauthoryear{Collett}{Collett}{2015}]{Collett:2015roa}
Collett T.~E.,  2015, \mn@doi [Astrophys. J.] {10.1088/0004-637X/811/1/20}, 811, 20

\bibitem[\protect\citeauthoryear{{Cranmer}, {Pavez}  \& {Louppe}}{{Cranmer} et~al.}{2015}]{cranmer2015approximating}
{Cranmer} K.,  {Pavez} J.,   {Louppe} G.,  2015, arXiv e-prints, \href {https://ui.adsabs.harvard.edu/abs/2015arXiv150602169C} {p. arXiv:1506.02169}

\bibitem[\protect\citeauthoryear{Cyr-Racine, Moustakas, Keeton, Sigurdson  \& Gilman}{Cyr-Racine et~al.}{2016}]{cyr-Racine2015darkcensus}
Cyr-Racine F.-Y.,  Moustakas L.~A.,  Keeton C.~R.,  Sigurdson K.,   Gilman D.~A.,  2016, \mn@doi [Phys. Rev. D] {10.1103/PhysRevD.94.043505}, 94, 043505

\bibitem[\protect\citeauthoryear{{Cyr-Racine}, {Keeton}  \& {Moustakas}}{{Cyr-Racine} et~al.}{2019}]{cyrracine2019}
{Cyr-Racine} F.-Y.,  {Keeton} C.~R.,   {Moustakas} L.~A.,  2019, \mn@doi [\prd] {10.1103/PhysRevD.100.023013}, \href {https://ui.adsabs.harvard.edu/abs/2019PhRvD.100b3013C} {100, 023013}

\bibitem[\protect\citeauthoryear{Dalal \& Kochanek}{Dalal \& Kochanek}{2002}]{dalal2001direction}
Dalal N.,  Kochanek C.~S.,  2002, \mn@doi [Astrophys. J.] {10.1086/340303}, 572, 25

\bibitem[\protect\citeauthoryear{{Daylan}, {Cyr-Racine}, {Diaz Rivero}, {Dvorkin}  \& {Finkbeiner}}{{Daylan} et~al.}{2018}]{daylan2018probing}
{Daylan} T.,  {Cyr-Racine} F.-Y.,  {Diaz Rivero} A.,  {Dvorkin} C.,   {Finkbeiner} D.~P.,  2018, \mn@doi [\apj] {10.3847/1538-4357/aaaa1e}, \href {https://ui.adsabs.harvard.edu/abs/2018ApJ...854..141D} {854, 141}

\bibitem[\protect\citeauthoryear{D\'\i{}az~Rivero, Cyr-Racine  \& Dvorkin}{D\'\i{}az~Rivero et~al.}{2018a}]{diazRivero2017powerspec}
D\'\i{}az~Rivero A.,  Cyr-Racine F.-Y.,   Dvorkin C.,  2018a, \mn@doi [Phys. Rev. D] {10.1103/PhysRevD.97.023001}, 97, 023001

\bibitem[\protect\citeauthoryear{D\'\i{}az~Rivero, Dvorkin, Cyr-Racine, Zavala  \& Vogelsberger}{D\'\i{}az~Rivero et~al.}{2018b}]{diazRivero2018powerspecethos}
D\'\i{}az~Rivero A.,  Dvorkin C.,  Cyr-Racine F.-Y.,  Zavala J.,   Vogelsberger M.,  2018b, \mn@doi [Phys. Rev. D] {10.1103/PhysRevD.98.103517}, 98, 103517

\bibitem[\protect\citeauthoryear{{Diemand} \& {Moore}}{{Diemand} \& {Moore}}{2011}]{diemand2011}
{Diemand} J.,  {Moore} B.,  2011, \mn@doi [Advanced Science Letters] {10.1166/asl.2011.1211}, \href {https://ui.adsabs.harvard.edu/abs/2011ASL.....4..297D} {4, 297}

\bibitem[\protect\citeauthoryear{{Dutton} \& {Macci{\`o}}}{{Dutton} \& {Macci{\`o}}}{2014}]{dutton2014cdmhalo}
{Dutton} A.~A.,  {Macci{\`o}} A.~V.,  2014, \mn@doi [\mnras] {10.1093/mnras/stu742}, \href {https://ui.adsabs.harvard.edu/abs/2014MNRAS.441.3359D} {441, 3359}

\bibitem[\protect\citeauthoryear{{Elbert}, {Bullock}, {Garrison-Kimmel}, {Rocha}, {O{\~n}orbe}  \& {Peter}}{{Elbert} et~al.}{2015}]{elbert2015core}
{Elbert} O.~D.,  {Bullock} J.~S.,  {Garrison-Kimmel} S.,  {Rocha} M.,  {O{\~n}orbe} J.,   {Peter} A. H.~G.,  2015, \mn@doi [\mnras] {10.1093/mnras/stv1470}, \href {https://ui.adsabs.harvard.edu/abs/2015MNRAS.453...29E} {453, 29}

\bibitem[\protect\citeauthoryear{{Fitts} et~al.,}{{Fitts} et~al.}{2017}]{fitts2017fire}
{Fitts} A.,  et~al., 2017, \mn@doi [\mnras] {10.1093/mnras/stx1757}, \href {https://ui.adsabs.harvard.edu/abs/2017MNRAS.471.3547F} {471, 3547}

\bibitem[\protect\citeauthoryear{{Gilman}, {Birrer}, {Treu}, {Keeton}  \& {Nierenberg}}{{Gilman} et~al.}{2018}]{gilman2018}
{Gilman} D.,  {Birrer} S.,  {Treu} T.,  {Keeton} C.~R.,   {Nierenberg} A.,  2018, \mn@doi [\mnras] {10.1093/mnras/sty2261}, \href {https://ui.adsabs.harvard.edu/abs/2018MNRAS.481..819G} {481, 819}

\bibitem[\protect\citeauthoryear{{Green}, {van den Bosch}  \& {Jiang}}{{Green} et~al.}{2021}]{green2021}
{Green} S.~B.,  {van den Bosch} F.~C.,   {Jiang} F.,  2021, \mn@doi [\mnras] {10.1093/mnras/stab696}, \href {https://ui.adsabs.harvard.edu/abs/2021MNRAS.503.4075G} {503, 4075}

\bibitem[\protect\citeauthoryear{{Hayashi}, {Navarro}, {Taylor}, {Stadel}  \& {Quinn}}{{Hayashi} et~al.}{2003}]{hayashi2003}
{Hayashi} E.,  {Navarro} J.~F.,  {Taylor} J.~E.,  {Stadel} J.,   {Quinn} T.,  2003, \mn@doi [\apj] {10.1086/345788}, \href {https://ui.adsabs.harvard.edu/abs/2003ApJ...584..541H} {584, 541}

\bibitem[\protect\citeauthoryear{He, Zhang, Ren  \& Sun}{He et~al.}{2016}]{he2016deepresiduallearning}
He K.,  Zhang X.,  Ren S.,   Sun J.,  2016, in 2016 IEEE Conference on Computer Vision and Pattern Recognition (CVPR). pp 770--778, \mn@doi{10.1109/CVPR.2016.90}

\bibitem[\protect\citeauthoryear{{He} et~al.,}{{He} et~al.}{2022}]{he2022}
{He} Q.,  et~al., 2022, \mn@doi [\mnras] {10.1093/mnras/stac191}, \href {https://ui.adsabs.harvard.edu/abs/2022MNRAS.511.3046H} {511, 3046}

\bibitem[\protect\citeauthoryear{{Hermans}, {Begy}  \& {Louppe}}{{Hermans} et~al.}{2019}]{hermans2019likelihoodfree}
{Hermans} J.,  {Begy} V.,   {Louppe} G.,  2019, arXiv e-prints, \href {https://ui.adsabs.harvard.edu/abs/2019arXiv190304057H} {p. arXiv:1903.04057}

\bibitem[\protect\citeauthoryear{Hezaveh, Dalal, Holder, Kisner, Kuhlen  \& Perreault~Levasseur}{Hezaveh et~al.}{2016a}]{hezaveh2014measuring}
Hezaveh Y.,  Dalal N.,  Holder G.,  Kisner T.,  Kuhlen M.,   Perreault~Levasseur L.,  2016a, \mn@doi [JCAP] {10.1088/1475-7516/2016/11/048}, 11, 048

\bibitem[\protect\citeauthoryear{{Hezaveh} et~al.,}{{Hezaveh} et~al.}{2016b}]{hezaveh2016detection}
{Hezaveh} Y.~D.,  et~al., 2016b, \mn@doi [\apj] {10.3847/0004-637X/823/1/37}, \href {https://ui.adsabs.harvard.edu/abs/2016ApJ...823...37H} {823, 37}

\bibitem[\protect\citeauthoryear{{Huang} et~al.,}{{Huang} et~al.}{2021}]{huang2021desi}
{Huang} X.,  et~al., 2021, \mn@doi [\apj] {10.3847/1538-4357/abd62b}, \href {https://ui.adsabs.harvard.edu/abs/2021ApJ...909...27H} {909, 27}

\bibitem[\protect\citeauthoryear{{Jacobs} et~al.,}{{Jacobs} et~al.}{2019}]{jacobs2019des}
{Jacobs} C.,  et~al., 2019, \mn@doi [\apjs] {10.3847/1538-4365/ab26b6}, \href {https://ui.adsabs.harvard.edu/abs/2019ApJS..243...17J} {243, 17}

\bibitem[\protect\citeauthoryear{{Kahlhoefer}, {Kaplinghat}, {Slatyer}  \& {Wu}}{{Kahlhoefer} et~al.}{2019}]{kahlhoefer2019sidm}
{Kahlhoefer} F.,  {Kaplinghat} M.,  {Slatyer} T.~R.,   {Wu} C.-L.,  2019, \mn@doi [\jcap] {10.1088/1475-7516/2019/12/010}, \href {https://ui.adsabs.harvard.edu/abs/2019JCAP...12..010K} {2019, 010}

\bibitem[\protect\citeauthoryear{Keeton, Kochanek  \& Seljak}{Keeton et~al.}{1997}]{Keeton1996shear}
Keeton C.~R.,  Kochanek C.~S.,   Seljak U.,  1997, \mn@doi [Astrophys. J.] {10.1086/304172}, 482, 604

\bibitem[\protect\citeauthoryear{{Kim}, {Peter}  \& {Hargis}}{{Kim} et~al.}{2018}]{kim2018missing}
{Kim} S.~Y.,  {Peter} A. H.~G.,   {Hargis} J.~R.,  2018, \mn@doi [\prl] {10.1103/PhysRevLett.121.211302}, \href {https://ui.adsabs.harvard.edu/abs/2018PhRvL.121u1302K} {121, 211302}

\bibitem[\protect\citeauthoryear{Kingma \& Ba}{Kingma \& Ba}{2014}]{kingma2014adam}
Kingma D.~P.,  Ba J.,  2014, arXiv preprint arXiv:1412.6980

\bibitem[\protect\citeauthoryear{{Kochanek} \& {White}}{{Kochanek} \& {White}}{2000}]{2000kochanek}
{Kochanek} C.~S.,  {White} M.,  2000, \mn@doi [\apj] {10.1086/317149}, \href {https://ui.adsabs.harvard.edu/abs/2000ApJ...543..514K} {543, 514}

\bibitem[\protect\citeauthoryear{{Koekemoer} et~al.,}{{Koekemoer} et~al.}{2007}]{koekemoer2007cosmos}
{Koekemoer} A.~M.,  et~al., 2007, \mn@doi [\apjs] {10.1086/520086}, \href {https://ui.adsabs.harvard.edu/abs/2007ApJS..172..196K} {172, 196}

\bibitem[\protect\citeauthoryear{{Laureijs} et~al.,}{{Laureijs} et~al.}{2011}]{laureijs2011euclid}
{Laureijs} R.,  et~al., 2011, arXiv e-prints, \href {https://ui.adsabs.harvard.edu/abs/2011arXiv1110.3193L} {p. arXiv:1110.3193}

\bibitem[\protect\citeauthoryear{{Loshchilov} \& {Hutter}}{{Loshchilov} \& {Hutter}}{2017}]{loshchilove2017adamw}
{Loshchilov} I.,  {Hutter} F.,  2017, arXiv e-prints, \href {https://ui.adsabs.harvard.edu/abs/2017arXiv171105101L} {p. arXiv:1711.05101}

\bibitem[\protect\citeauthoryear{{Lovell} et~al.,}{{Lovell} et~al.}{2012}]{lovell2012wdm}
{Lovell} M.~R.,  et~al., 2012, \mn@doi [\mnras] {10.1111/j.1365-2966.2011.20200.x}, \href {https://ui.adsabs.harvard.edu/abs/2012MNRAS.420.2318L} {420, 2318}

\bibitem[\protect\citeauthoryear{{Lovell}, {Frenk}, {Eke}, {Jenkins}, {Gao}  \& {Theuns}}{{Lovell} et~al.}{2014}]{lovell2014wdm}
{Lovell} M.~R.,  {Frenk} C.~S.,  {Eke} V.~R.,  {Jenkins} A.,  {Gao} L.,   {Theuns} T.,  2014, \mn@doi [\mnras] {10.1093/mnras/stt2431}, \href {https://ui.adsabs.harvard.edu/abs/2014MNRAS.439..300L} {439, 300}

\bibitem[\protect\citeauthoryear{{Lynden-Bell} \& {Wood}}{{Lynden-Bell} \& {Wood}}{1968}]{lyndenbell1968}
{Lynden-Bell} D.,  {Wood} R.,  1968, \mn@doi [\mnras] {10.1093/mnras/138.4.495}, \href {https://ui.adsabs.harvard.edu/abs/1968MNRAS.138..495L} {138, 495}

\bibitem[\protect\citeauthoryear{{Mandelbaum}, {Hirata}, {Leauthaud}, {Massey}  \& {Rhodes}}{{Mandelbaum} et~al.}{2012}]{mandelbaum2012precision}
{Mandelbaum} R.,  {Hirata} C.~M.,  {Leauthaud} A.,  {Massey} R.~J.,   {Rhodes} J.,  2012, \mn@doi [\mnras] {10.1111/j.1365-2966.2011.20138.x}, \href {https://ui.adsabs.harvard.edu/abs/2012MNRAS.420.1518M} {420, 1518}

\bibitem[\protect\citeauthoryear{{Mandelbaum} et~al.,}{{Mandelbaum} et~al.}{2014}]{mandelbaum2014great3}
{Mandelbaum} R.,  et~al., 2014, \mn@doi [\apjs] {10.1088/0067-0049/212/1/5}, \href {https://ui.adsabs.harvard.edu/abs/2014ApJS..212....5M} {212, 5}

\bibitem[\protect\citeauthoryear{{McKean} et~al.,}{{McKean} et~al.}{2015}]{mckean2015ska}
{McKean} J.,  et~al., 2015, in Advancing Astrophysics with the Square Kilometre Array (AASKA14). p.~84 (\mn@eprint {arXiv} {1502.03362})

\bibitem[\protect\citeauthoryear{{Meneghetti} et~al.,}{{Meneghetti} et~al.}{2020}]{meneghetti2020}
{Meneghetti} M.,  et~al., 2020, \mn@doi [Science] {10.1126/science.aax5164}, \href {https://ui.adsabs.harvard.edu/abs/2020Sci...369.1347M} {369, 1347}

\bibitem[\protect\citeauthoryear{{Meneghetti} et~al.,}{{Meneghetti} et~al.}{2022}]{meneghetti2022}
{Meneghetti} M.,  et~al., 2022, \mn@doi [\aap] {10.1051/0004-6361/202243779}, \href {https://ui.adsabs.harvard.edu/abs/2022A&A...668A.188M} {668, A188}

\bibitem[\protect\citeauthoryear{{Meneghetti} et~al.,}{{Meneghetti} et~al.}{2023}]{meneghetti2023}
{Meneghetti} M.,  et~al., 2023, \mn@doi [\aap] {10.1051/0004-6361/202346975}, \href {https://ui.adsabs.harvard.edu/abs/2023A&A...678L...2M} {678, L2}

\bibitem[\protect\citeauthoryear{{Minor}, {Kaplinghat}, {Chan}  \& {Simon}}{{Minor} et~al.}{2021a}]{minor2021inferring}
{Minor} Q.,  {Kaplinghat} M.,  {Chan} T.~H.,   {Simon} E.,  2021a, \mn@doi [\mnras] {10.1093/mnras/stab2209}, \href {https://ui.adsabs.harvard.edu/abs/2021MNRAS.507.1202M} {507, 1202}

\bibitem[\protect\citeauthoryear{{Minor}, {Gad-Nasr}, {Kaplinghat}  \& {Vegetti}}{{Minor} et~al.}{2021b}]{minor2021unexpected}
{Minor} Q.,  {Gad-Nasr} S.,  {Kaplinghat} M.,   {Vegetti} S.,  2021b, \mn@doi [\mnras] {10.1093/mnras/stab2247}, \href {https://ui.adsabs.harvard.edu/abs/2021MNRAS.507.1662M} {507, 1662}

\bibitem[\protect\citeauthoryear{{Mohamed} \& {Lakshminarayanan}}{{Mohamed} \& {Lakshminarayanan}}{2016}]{mohamed2016learning}
{Mohamed} S.,  {Lakshminarayanan} B.,  2016, arXiv e-prints, \href {https://ui.adsabs.harvard.edu/abs/2016arXiv161003483M} {p. arXiv:1610.03483}

\bibitem[\protect\citeauthoryear{{Nadler}, {Yang}  \& {Yu}}{{Nadler} et~al.}{2023}]{nadler2023}
{Nadler} E.~O.,  {Yang} D.,   {Yu} H.-B.,  2023, \mn@doi [arXiv e-prints] {10.48550/arXiv.2306.01830}, \href {https://ui.adsabs.harvard.edu/abs/2023arXiv230601830N} {p. arXiv:2306.01830}

\bibitem[\protect\citeauthoryear{{Navarro}, {Frenk}  \& {White}}{{Navarro} et~al.}{1997}]{navarro1997nfw}
{Navarro} J.~F.,  {Frenk} C.~S.,   {White} S. D.~M.,  1997, \mn@doi [\apj] {10.1086/304888}, \href {https://ui.adsabs.harvard.edu/abs/1997ApJ...490..493N} {490, 493}

\bibitem[\protect\citeauthoryear{{Nishikawa}, {Boddy}  \& {Kaplinghat}}{{Nishikawa} et~al.}{2020}]{nishikawa2020}
{Nishikawa} H.,  {Boddy} K.~K.,   {Kaplinghat} M.,  2020, \mn@doi [\prd] {10.1103/PhysRevD.101.063009}, \href {https://ui.adsabs.harvard.edu/abs/2020PhRvD.101f3009N} {101, 063009}

\bibitem[\protect\citeauthoryear{Ostdiek, Diaz~Rivero  \& Dvorkin}{Ostdiek et~al.}{2022a}]{Ostdiek:2020cqz}
Ostdiek B.,  Diaz~Rivero A.,   Dvorkin C.,  2022a, \mn@doi [Astron. Astrophys.] {10.1051/0004-6361/202142030}, 657, L14

\bibitem[\protect\citeauthoryear{Ostdiek, Diaz~Rivero  \& Dvorkin}{Ostdiek et~al.}{2022b}]{Ostdiek:2020mvo}
Ostdiek B.,  Diaz~Rivero A.,   Dvorkin C.,  2022b, \mn@doi [Astrophys. J.] {10.3847/1538-4357/ac2d8d}, 927, 83

\bibitem[\protect\citeauthoryear{Paszke et~al.,}{Paszke et~al.}{2019}]{NEURIPS2019_9015}
Paszke A.,  et~al., 2019, in Wallach H.,  Larochelle H.,  Beygelzimer A.,  d\textquotesingle Alch\'{e}-Buc F.,  Fox E.,   Garnett R.,  eds, , Advances in Neural Information Processing Systems 32.
Curran Associates, Inc., pp 8024--8035, \url {http://papers.neurips.cc/paper/9015-pytorch-an-imperative-style-high-performance-deep-learning-library.pdf}

\bibitem[\protect\citeauthoryear{{Read}, {Iorio}, {Agertz}  \& {Fraternali}}{{Read} et~al.}{2017}]{read2017stellar}
{Read} J.~I.,  {Iorio} G.,  {Agertz} O.,   {Fraternali} F.,  2017, \mn@doi [\mnras] {10.1093/mnras/stx147}, \href {https://ui.adsabs.harvard.edu/abs/2017MNRAS.467.2019R} {467, 2019}

\bibitem[\protect\citeauthoryear{{Recht}, {Roelofs}, {Schmidt}  \& {Shankar}}{{Recht} et~al.}{2018}]{recht2018}
{Recht} B.,  {Roelofs} R.,  {Schmidt} L.,   {Shankar} V.,  2018, \mn@doi [arXiv e-prints] {10.48550/arXiv.1806.00451}, \href {https://ui.adsabs.harvard.edu/abs/2018arXiv180600451R} {p. arXiv:1806.00451}

\bibitem[\protect\citeauthoryear{{Recht}, {Roelofs}, {Schmidt}  \& {Shankar}}{{Recht} et~al.}{2019}]{recht2019}
{Recht} B.,  {Roelofs} R.,  {Schmidt} L.,   {Shankar} V.,  2019, \mn@doi [arXiv e-prints] {10.48550/arXiv.1902.10811}, \href {https://ui.adsabs.harvard.edu/abs/2019arXiv190210811R} {p. arXiv:1902.10811}

\bibitem[\protect\citeauthoryear{{Ritondale}, {Vegetti}, {Despali}, {Auger}, {Koopmans}  \& {McKean}}{{Ritondale} et~al.}{2019}]{ritondale2019lowmass}
{Ritondale} E.,  {Vegetti} S.,  {Despali} G.,  {Auger} M.~W.,  {Koopmans} L.~V.~E.,   {McKean} J.~P.,  2019, \mn@doi [\mnras] {10.1093/mnras/stz464}, \href {https://ui.adsabs.harvard.edu/abs/2019MNRAS.485.2179R} {485, 2179}

\bibitem[\protect\citeauthoryear{{Rocha}, {Peter}, {Bullock}, {Kaplinghat}, {Garrison-Kimmel}, {O{\~n}orbe}  \& {Moustakas}}{{Rocha} et~al.}{2013}]{rocha2013sidm}
{Rocha} M.,  {Peter} A. H.~G.,  {Bullock} J.~S.,  {Kaplinghat} M.,  {Garrison-Kimmel} S.,  {O{\~n}orbe} J.,   {Moustakas} L.~A.,  2013, \mn@doi [\mnras] {10.1093/mnras/sts514}, \href {https://ui.adsabs.harvard.edu/abs/2013MNRAS.430...81R} {430, 81}

\bibitem[\protect\citeauthoryear{{Scoville} et~al.,}{{Scoville} et~al.}{2007}]{scoville2007cosmos}
{Scoville} N.,  et~al., 2007, \mn@doi [\apjs] {10.1086/516585}, \href {https://ui.adsabs.harvard.edu/abs/2007ApJS..172....1S} {172, 1}

\bibitem[\protect\citeauthoryear{{S{\'e}rsic}}{{S{\'e}rsic}}{1963}]{sersic1963}
{S{\'e}rsic} J.~L.,  1963, Boletin de la Asociacion Argentina de Astronomia La Plata Argentina, \href {https://ui.adsabs.harvard.edu/abs/1963BAAA....6...41S} {6, 41}

\bibitem[\protect\citeauthoryear{Sheth, Mo  \& Tormen}{Sheth et~al.}{2001}]{sheth2001}
Sheth R.~K.,  Mo H.~J.,   Tormen G.,  2001, \mn@doi [Monthly Notices of the Royal Astronomical Society] {10.1046/j.1365-8711.2001.04006.x}, 323, 1

\bibitem[\protect\citeauthoryear{{Spergel} \& {Steinhardt}}{{Spergel} \& {Steinhardt}}{2000}]{spergel2000sidm}
{Spergel} D.~N.,  {Steinhardt} P.~J.,  2000, \mn@doi [\prl] {10.1103/PhysRevLett.84.3760}, \href {https://ui.adsabs.harvard.edu/abs/2000PhRvL..84.3760S} {84, 3760}

\bibitem[\protect\citeauthoryear{{Storfer} et~al.,}{{Storfer} et~al.}{2022}]{storfer2022desi}
{Storfer} C.,  et~al., 2022, arXiv e-prints, \href {https://ui.adsabs.harvard.edu/abs/2022arXiv220602764S} {p. arXiv:2206.02764}

\bibitem[\protect\citeauthoryear{{Vegetti}, {Koopmans}, {Bolton}, {Treu}  \& {Gavazzi}}{{Vegetti} et~al.}{2010}]{vegetti2010detection}
{Vegetti} S.,  {Koopmans} L.~V.~E.,  {Bolton} A.,  {Treu} T.,   {Gavazzi} R.,  2010, \mn@doi [\mnras] {10.1111/j.1365-2966.2010.16865.x}, \href {https://ui.adsabs.harvard.edu/abs/2010MNRAS.408.1969V} {408, 1969}

\bibitem[\protect\citeauthoryear{{Vegetti}, {Lagattuta}, {McKean}, {Auger}, {Fassnacht}  \& {Koopmans}}{{Vegetti} et~al.}{2012}]{vegetti2012detection}
{Vegetti} S.,  {Lagattuta} D.~J.,  {McKean} J.~P.,  {Auger} M.~W.,  {Fassnacht} C.~D.,   {Koopmans} L.~V.~E.,  2012, \mn@doi [\nat] {10.1038/nature10669}, \href {https://ui.adsabs.harvard.edu/abs/2012Natur.481..341V} {481, 341}

\bibitem[\protect\citeauthoryear{{Vegetti}, {Koopmans}, {Auger}, {Treu}  \& {Bolton}}{{Vegetti} et~al.}{2014}]{vegetti2014inferencecdmsubmf}
{Vegetti} S.,  {Koopmans} L.~V.~E.,  {Auger} M.~W.,  {Treu} T.,   {Bolton} A.~S.,  2014, \mn@doi [\mnras] {10.1093/mnras/stu943}, \href {https://ui.adsabs.harvard.edu/abs/2014MNRAS.442.2017V} {442, 2017}

\bibitem[\protect\citeauthoryear{{Vogelsberger}, {Zavala}  \& {Loeb}}{{Vogelsberger} et~al.}{2012}]{vogelsberger2012sidm}
{Vogelsberger} M.,  {Zavala} J.,   {Loeb} A.,  2012, \mn@doi [\mnras] {10.1111/j.1365-2966.2012.21182.x}, \href {https://ui.adsabs.harvard.edu/abs/2012MNRAS.423.3740V} {423, 3740}

\bibitem[\protect\citeauthoryear{{Wagner-Carena}, {Aalbers}, {Birrer}, {Nadler}, {Darragh-Ford}, {Marshall}  \& {Wechsler}}{{Wagner-Carena} et~al.}{2023}]{wagner2022images}
{Wagner-Carena} S.,  {Aalbers} J.,  {Birrer} S.,  {Nadler} E.~O.,  {Darragh-Ford} E.,  {Marshall} P.~J.,   {Wechsler} R.~H.,  2023, \mn@doi [\apj] {10.3847/1538-4357/aca525}, \href {https://ui.adsabs.harvard.edu/abs/2023ApJ...942...75W} {942, 75}

\bibitem[\protect\citeauthoryear{Wilks}{Wilks}{1938}]{Wilks:1938dza}
Wilks S.~S.,  1938, \mn@doi [Annals Math. Statist.] {10.1214/aoms/1177732360}, 9, 60

\bibitem[\protect\citeauthoryear{Yang \& Yu}{Yang \& Yu}{2021}]{core_collapse}
Yang D.,  Yu H.-B.,  2021, \mn@doi [Phys. Rev. D] {10.1103/PhysRevD.104.103031}, 104, 103031

\bibitem[\protect\citeauthoryear{Zeng, Peter, Du, Benson, Kim, Jiang, Cyr-Racine  \& Vogelsberger}{Zeng et~al.}{2022}]{core_collapse_hybrid}
Zeng Z.~C.,  Peter A. H.~G.,  Du X.,  Benson A.,  Kim S.,  Jiang F.,  Cyr-Racine F.-Y.,   Vogelsberger M.,  2022, \mn@doi [Monthly Notices of the Royal Astronomical Society] {10.1093/mnras/stac1094}, 513, 4845

\bibitem[\protect\citeauthoryear{{Zhang}, {Mishra-Sharma}  \& {Dvorkin}}{{Zhang} et~al.}{2022}]{zhang2022}
{Zhang} G.,  {Mishra-Sharma} S.,   {Dvorkin} C.,  2022, \mn@doi [\mnras] {10.1093/mnras/stac3014}, \href {https://ui.adsabs.harvard.edu/abs/2022MNRAS.517.4317Z} {517, 4317}

\bibitem[\protect\citeauthoryear{{{\c{S}}eng{\"u}l} \& {Dvorkin}}{{{\c{S}}eng{\"u}l} \& {Dvorkin}}{2022}]{sengul2022probing}
{{\c{S}}eng{\"u}l} A.~{\c{C}}.,  {Dvorkin} C.,  2022, \mn@doi [\mnras] {10.1093/mnras/stac2256}, \href {https://ui.adsabs.harvard.edu/abs/2022MNRAS.516..336S} {516, 336}

\bibitem[\protect\citeauthoryear{{{\c{S}}eng{\"u}l}, {Dvorkin}, {Ostdiek}  \& {Tsang}}{{{\c{S}}eng{\"u}l} et~al.}{2022}]{sengul2022reanalyzed}
{{\c{S}}eng{\"u}l} A.~{\c{C}}.,  {Dvorkin} C.,  {Ostdiek} B.,   {Tsang} A.,  2022, \mn@doi [\mnras] {10.1093/mnras/stac1967}, \href {https://ui.adsabs.harvard.edu/abs/2022MNRAS.515.4391S} {515, 4391}

\makeatother
\end{thebibliography}

\end{document}